\pdfoutput=1
\documentclass[11pt,twoside]{article}
\usepackage[backend=biber]{biblatex}
\bibliography{Dynamic_Estimates_Of_The_Arrow_Pratt_Absolute_And_Relative_Risk_Aversion_Coefficients.bib}
\usepackage[utf8]{inputenc}
\usepackage{mathtools}
\usepackage{etoolbox}
\usepackage{enumitem}
\usepackage{adjustbox}
\usepackage{graphics}
\usepackage{fancyhdr}
\usepackage[toc,page]{appendix}
\makeatletter
\usepackage{ctable}
\usepackage[flushleft]{threeparttable}
\usepackage{titletoc}
\usepackage{fix-cm}
\usepackage{datetime}
\usepackage{ragged2e}
\usepackage{lipsum}
\usepackage{subcaption}
\usepackage{float}
\usepackage{footnote}
\usepackage{amsmath, amssymb, amsthm} % AMS packages
\usepackage{graphicx,color}           % Packages for graphics and color
\usepackage[left=1in, right=1in, top=1in, bottom=1in, includefoot]{geometry}
\usepackage{booktabs,caption}
\usepackage{dsfont}
\usepackage{multirow,array}
\usepackage{xcolor}
\usepackage{titlesec}
\newtheorem{theorem}{Theorem}[section]
\numberwithin{theorem}{section}

\newtheorem{definition}[theorem]{Definition}

\DeclareFieldFormat[article, inbook]{title}{#1}

\title{Dynamic Estimates Of The Arrow-Pratt Absolute And Relative Risk Aversion Coefficients}

\date{\today}

\author{ George Samartzis\\
	Department of Banking and Financial Management\\
	University of Piraeus\\
	Piraeus, 18534 \\
	\texttt{gsamartzis@unipi.gr} \\
	\\
	
	Nikitas Pittis\\
	Department of Banking and Financial Management\\
	University of Piraeus\\
	Piraeus, 18534 \\
	\texttt{npittis@unipi.gr} \\
}

\providecommand{\keywords}[1]{\textbf{\textit{Index terms }} #1}

\begin{document}
	\maketitle
	
	\begin{abstract}
		We derive a closed-form expression capturing the degree of Relative Risk Aversion (RRA) of investors for non-"fair" lotteries. We argue that our formula is superior to earlier methods that have been proposed, as it is a function of only three variables. Namely, the Treasury yields, the returns and the market capitalization of a specific market index. Our formula, is tested on CAC 40, EURO, S\&P 500 and STOXX 600, with respect to the market capitalization of each index, for different time periods. We deduce that the investors in these markets exhibit Decreasing Absolute Risk Aversion (DARA) through all the different time periods that we consider, while the degree of RRA has altered between being constant, decreasing or increasing. Furthermore, we propose a simple and intuitive way to measure the degree to which a wrong assumption with respect to the utility function of an investor will affect the structure of his portfolio. Our method is built on a two asset portfolio framework. Namely, a portfolio consisting of one risky and one risk-free asset. Applying our method, the empirical findings indicate that the weight invested in the risky asset varies substantially even among utility functions with similar characteristics.
	\end{abstract}

	% keywords can be removed
	\keywords{ARA, \and RRA}

	\section{Introduction}
	The risk attitude of investors is considered to be a crucial factor in Portfolio and Decision Theory. A rational investor in the Von Neumann-Morgenstern sense \cite{Neumann_Morgenstern}, will always choose the lottery that maximizes his/her expected utility function. Based on that decision criterion, Arrow and Pratt \cite{Arrow1965}, \cite{Pratt1964} developed the concepts of absolute and relative risk aversion (hereafter; ARA and RRA) as ways to indicate the level of risk-aversion of the investor. Determining the trends of ARA and RRA is crucial for pointing at a specific class of utility functions.
	
	The existing literature has done a pile of work on the recovery of degree of ARA and RRA of investors. In terms of ARA, the literature universally finds evidence of decreasing ARA (DARA) as wealth increases. This is in agreement with Arrow \cite{Arrow1965} who claimed that the DARA "seems supported by everyday observation". On the other hand, the empirical works on RRA derive mixed conclusions. Some of the earliest studies that found evidence of DRRA with respect to wealth are \cite{Cohn1975}, \cite{Morin1983}, \cite{Levy1994}, \cite{Guiso1996}, \cite{Ogaki2001} and \cite{Perignon2002}. While evidence of increasing RRA (IRRA) was found in \cite{Siegel1982}, \cite{Barsky1997}, \cite{Bar-Shira1997} and \cite{Eisenhauer2003}. Most of these works, based their findings on cross-section data, which implicitly assumes that all investors have the same utility function, a rather unrealistic assumption. Works like that of \cite{Levy1994}, \cite{Barsky1997}, \cite{Bar-Shira1997} and \cite{Eisenhauer2003} used both experimental and econometric methods to extract RRA. A totally different approach was used by \cite{Jackwerth2000} and \cite{Perignon2002} who utilized information from Options markets. The most recent part of the literature, argues that investors reveal constant RRA (CRRA). Some of the works supporting this evidence are \cite{Sahm2012}, \cite{Brunnermeier2008} and \cite{Chiappori2011}. The advantage of these studies is that they all use panel data. So, the implicit assumption of a common utility function among investors is no longer made.
	
	Our work, revolves around the definitions of ARA and RRA and the concept of a lottery that is not "fair". Considering the fact that markets are not "fair" lotteries the value of such an approach becomes evident. We proceed with deriving the formulae of the absolute and relative risk premium which subsequently lead to ARA and RRA when the underlying lottery is not "fair". These formulae can be used to extract the risk attitude of investors across different markets, specifically, in terms of ARA and RRA.
	
	Using monthly returns and market cap from CAC 40, EURO, S\&P 500 and STOXX 600 as well as 10-year Treasury yields, we find evidence of DARA and IRRA for 2012 to 2022, for all markets. Moreover, our results, capture the impact of the QE program announced by Fed in March 2020. We also test our findings using a rolling-window approach with 5 and 15-year window sizes to assess their robustness. The new results do not deviate from our prior conclusions. Then, we use data for different time periods, namely, 1993-1998 and 1999-2005 to determine whether or not the investors' risk attitude is consistent through years. Our findings, point at DARA and DRRA for 1993-1998 and CRRA/IRRA and DARA for 1999-2005. As a result, we conclude that through the years the markets became more risk-averse with respect to their level of wealth. 
	
	The evidence of DARA and IRRA for 2012-2022 is satisfied by a specific class of utility functions. This leads us to our next question which has to do with measuring the impact of a wrong assumption with respect to the utility function of an investor. For that, we consider an asset manager who mistakenly assumes that his customer is characterized by a quadratic utility function while in reality the customer's true utility function is logarithmic. Assuming that the asset manager diversifies the investor's wealth between a risky and a risk-free asset we propose a way to measure the differences in the asset manager's decisions on behalf of the investor for different utility functions. The results predicate that the asset manager will mislead his customer considerably, in case he assumes quadratic instead of logarithmic utility functions.
	
	Our main findings split into four parts. (i) We derive a closed-form expression for the degree of RRA for a lottery with nonzero mean. (ii) We apply our formula on four different markets, namely CAC 40, EURO, S\&P 500 and STOXX 600, to deduce that all of them exhibit DARA and IRRA. (iii) We also find evidence that the slope of the RRA has changed through different time periods, while the slope of ARA has always been decreasing. (iv) Finally, we introduce a way to measure the differences in portfolio diversification among different utility functions.
	
	\section{Theoretical Framework}
	\subsection{Degree Of Risk Aversion For A "Fair" Lottery}
	In this section, we delve into the different notions underlying Decision Theory. Consider the following case. A decision maker (or investor) with level of wealth $w_0$ is asked whether or not he wants to participate in a game (or lottery) represented by $Z$. Assuming that his binary relation ($\succeq$) satisfies the axioms posed by VN-M for any lottery, we can safely say that his decision criterion arises from the Representation Theorem of VN-M. Namely, he should always choose the lottery that maximizes his expected utility function. Now, in order to participate in the game the investor demands a risk premium represented by $\rho(w_0,Z)$. The risk premium is based on the notion of Certainty Equivalent (hereafter; CE). The CE is defined below.
	\begin{definition}
		The \textbf{Certainty Equivalent} of lottery $Z$ is the amount $z_0$ which makes the decision maker indifferent between lottery $Z$ and this certain amount $z_0$. Alternatively, if the decision maker owns lottery $Z$, then $z_0$ represents the minimum amount he is willing to sell the lottery. Mathematically,
		\begin{equation*}
			\zeta_{z_0}\sim Z\overset{VN-M}{\Leftrightarrow}U(z_0)=E[U(Z)]\Leftrightarrow z_0=U^{-1}(E[U(Z)]), \quad \forall Z
		\end{equation*}
	\end{definition}
	\noindent
	From the definition, we see that the CE results from the decision maker's subjective characteristics, but is also dependent on the objective (probabilistic) characteristics of lottery $Z$. Thus, the level of "fear" the decision maker has with respect to risk will determine the magnitude of CE for bearing the risk of $Z$. This leads to the definition of risk premium.
	\begin{definition}
		The \textbf{risk premium} is the difference between the expected return of lottery $Z$ and the Certainty Equivalent the decision maker asks for it.
		\begin{equation*}
			\rho(Z) = E[Z]-z_0=\mu_Z-z_0.
		\end{equation*}
	\end{definition}
	\noindent
	The notion of risk premium translates in two ways. (i) If the decision maker owns lottery $Z$, $\rho(Z)$ represents the expected units of return he is willing to "sacrifice" in order to avoid its risk. (ii) Alternatively, if the decision maker considers owning lottery $Z$, $\rho(Z)$ represents the expected units of return he "demands" in order to bear the risk of lottery $Z$. In our frame, we will focus exclusively in the case of risk-averse investors with an increasing utility function $U'>0$. That implies that the following should hold:
	\begin{itemize}
		\item The utility function $U$ is \textbf{concave}, i.e. $U''<0$
		\item The risk premium is positive $\rho(Z)>0, \ \forall Z$
	\end{itemize}
	
	Now, assume that on a later stage of his life the investor has a new higher level of wealth $w_0^+$ ($w_0^+>w_0$). What is the new risk premium $\rho(w_0^+,Z)$ the investor demands assuming that the objective characteristics of lottery $Z$ remain unchanged? Although, it can be the same we shall consider the case where
	\begin{equation*}
		\rho(w_0^+,Z)\neq \rho(w_0,Z).
	\end{equation*}
	\noindent
	Since the mean and variance of lottery $Z$ remain the same what is the causative factor for the change in $\rho$? There should be a function of $w_0$ which we will express by $r(w_0)$ that affects the risk premium. Based on Arrow and Pratt this function is called local degree of absolute risk aversion (hereafter; ARA) of the investor with respect to his level of wealth. This measure is defined below.
	\begin{definition}
		The local degree of \textbf{Absolute Risk Aversion}, $r(w)$, at a level of wealth $W$ is defined as
		\begin{equation*}
			r(W)=-\frac{U''(W)}{U'(W)}
		\end{equation*}
	\end{definition}
	\noindent
	Interpretation of ARA:
	\begin{itemize}
		\item For risk-averse investors we have $U''<0$. The larger $|U''|$ it is, the more concave the utility function of the investor will be. In other words, the investor will be more risk-averse.
		\item According to VN-M, any positive linear transformation of $U$ should still satisfy the Representation theorem and by extension lead to the same $r(W)$. That justifies the division by $U'$.
		\item ARA overcomes an important limitation of the risk premium. Namely, in order to compare the degree of risk-aversion of two investors, we would need to compare their risk premiums for any lottery $Z$ and any level of wealth $W$. On the contrary, $r(W)$ needs to be tested just for any level of wealth $W$.
		\item This degree of risk aversion is measured in absolute terms (dollars).
	\end{itemize}
	Now that we specified the factor that generates the differences in the aforementioned example, we will derive $\rho(w_0,Z)$ in terms of $r(w_0)$. In fact, we assume that $Z$ is a "fair" lottery in the sense that $\mu_Z=0$ and $\sigma_Z$ is relatively small. Let also, the final level of wealth of the decision maker expressed by
	\begin{equation*}
		W_1=w_0+Z.
	\end{equation*}
	Using the definition of CE together with 2nd degree Taylor series we derive
	\begin{equation}
		\rho(W_1)=\rho(w_0,Z)=\rho(w_0+Z)=\frac{1}{2}r(w_0)\sigma_Z^2.
	\end{equation}
	Interpretation of $\rho(w_0,Z)$:
	\begin{itemize}
		\item The subjective factor, $r(w_0)$, remains unchanged regardless of any changes in the objective characteristics of lottery $Z$. However, the risk premium varies for different lotteries. This is due to the objective factor of the lottery, $\sigma_Z^2$.
		\item If $r(w_0)$ is decreasing (DARA), increasing (IARA) or constant (CARA) for any level of wealth $w_0$, then $\rho(w_0,Z)$ will also be decreasing, increasing, or constant, respectively.
	\end{itemize}
	Within the current context, lottery $Z$ is independent of any change in $w_0$. Consider the following case. A decision maker with $w_0=1000\$$ is asked to participate in a coin toss game with the following set of outcomes. If the coin lands "Heads" the decision maker will earn $100\$$ while if it lands "Tails" he will have to pay $100\$$. The decision maker decides that this lottery is too risky for his risk appetite. Assume that the decision maker's degree of risk aversion is DARA. So, at a later stage of his life his wealth is $100,000\$$ and thus he feels comfortable to participate in exactly the same game. Now, assume that the same decision maker is asked to participate in a different coin toss game $Z'$ in which if the coin lands "Heads" he will earn $10\%$ of his wealth $w_0$ while if it lands "Tails" he will have to pay $10\%$ of $w_0$. That means that for $w_0=1000\$$ he will earn/lose $100\$$ while for $w_0=100,000\$$ earn/lose $10,000\$$. Obviously, this concept may alter the decision made by the decision maker. In particular, the fact that his wealth increased does not necessarily mean that he will participate in $Z'$ since this new game changes with his level of wealth, i.e. $Z'=w_0R$, where $R$ is a lottery expressed in percentages.
	
	To deal with such cases, Arrow and Pratt defined a new measure called the local degree of Relative Risk Aversion (hereafter; RRA).
	\begin{definition}
		The local degree of \textbf{Relative Risk Aversion}, $\lambda(w)$, at a level of wealth $W$ is defined as
		\begin{equation*}
			\lambda(W)=-W\frac{U''(W)}{U'(W)}=Wr(W)
		\end{equation*}
	\end{definition}
	\noindent
	Interpretation of RRA:
	\begin{itemize}
		\item RRA depends on ARA, since $\lambda(W)=Wr(W)$. If for example $r(W)$ is decreasing at a faster rate than the rate at which $W$ is increasing, $r(W)$ will also be a decreasing function of $W$ (DRRA). On the other side, if $r(W)$ is decreasing at a slower rate than the rate at which $W$ is increasing, $r(W)$ will be an increasing function of $W$ (IRRA).
		\item This degree of risk aversion is measured in relative terms (percentages).
	\end{itemize}
	Under this new measure, consider a "fair" lottery $Z=Rw_0$ with $\mu_Z=\mu_R=0$ and $\sigma_Z,\sigma_R$ is relatively small. Then, the "relative" risk premium for $W_1$ will be
	\begin{align}
		\begin{split}
			&\rho(W_1)=\rho(w_0,Rw_0)=\frac{1}{2}r(w_0)\sigma_{Rw_0}^2\\
			&\frac{\rho(w_0,Rw_0)}{w_0}=\frac{1}{2}\frac{w_0^2r(w_0)\sigma_R^2}{w_0}\\
			&\tilde{\rho}_{w_0}(R)=\frac{1}{2}\lambda(w_0)\sigma_R^2.
		\end{split}
	\end{align}
	Based on the above analysis, the literature has done extensive research on the Decision Making of investors under Risk. More specifically, as we will see in the following subsections the literature focused on identifying the slope of the ARA and the RRA of investors with respect to their level of wealth. We will see that previous research works have followed multiple approaches in trying to arrive at more reliable conclusions. Following our overview, we will introduce a totally unique approach on this subject.
	
	\subsection{Decreasing Absolute Risk Aversion Literature}
	In 1965, Arrow \cite{Arrow1965} posed the hypothesis that investors reveal DARA and IRRA. In fact, he claimed that the DARA "seems supported by everyday observation". However, in terms of IRRA Arrow said that: "the hypothesis of increasing relative risk aversion is not easily confrontable with intuitive evidence". From that point on, an abundance of research works have attempted to recover the relative risk preferences of the investors, while DARA is universally accepted. Extracting RRA, has been proven to be particularly challenging. We will review the research work on RRA in the subsequent sections. 
	
	Haim Levy (1994) \cite{Levy1994}, applying time-series analysis found that "only six subjects out of 62 significantly contradict the DARA property". As a matter of fact, Levy concluded that the evidence is significantly strong and supports Arrow's assertion. We will analyze Levy's approach in the next section where we review his findings on RRA.
	
	Some other studies that find supportive evidence of DARA are \cite{Bar-Shira1997}, \cite{Jackwerth2000} and \cite{Eisenhauer2003}. Going on, the literature considered DARA to be the norm and focused more on RRA. This universally accepted property of individual risk preferences plays a crucial role in many applications of the expected utility theory.
	
	\subsection{Decreasing Relative Risk Aversion Literature}
	Let us focus on the bibliography that shows evidence of DRRA. Cohn et al. \cite{Cohn1975} focused on the information from 588 different portfolio allocations provided by customers of a retail brokerage firm. In particular, each customer provided information with respect to the share he had in common stocks, corporate bonds, etc., in percentages. The authors established two alternative classifications for long-term fixed-income securities. In the first one, they treated Savings Account, Checking Account, Personal Residence, Personal Property and Other Assets as the "risk-free" assets. While, in the second Classification they also included Preferred Stock, Corporate Bonds and Government Bonds in this category. However, there is no compelling reason to treat Corporate Bonds as riskless assets. The level of wealth was also given two different definitions. With the first one being the Total Assets and the second one being Total Assets less Personal Residence. Although, they acknowledged the fact that a preferred proxy of wealth would be the Net Worth of  each customer, which was not possible as the customers did not provide any data on their Liabilities. Such an approach could conceivably alter their conclusions. The authors found strong evidence of DRRA even when they controlled for demographic factors like the age of each customer as well as his/her marital status. In their study, they split the customers into different wealth groups, in order to determine whether or not their findings are consistent between different groups. The fact that their target group pertains to active investors only, means that they do not cover a complete cross-section of investors at each wealth level. Finally, the aforementioned results depend on cross-sectional data which means that the authors did not study the portfolio allocation of each individual in time. This implies that they draw conclusions based on information from individuals with different utility functions. 
	
	Morin and Suarez \cite{Morin1983}, used portfolio allocation data of 9,962 different private households taken from the Survey of Consumer Finances (SCF) in Canada. The authors argue that the SCF database used is more representative of investors and broad-based in terms of range of wealth covered. One major improvement in their methodology is that they use Net worth as a proxy of wealth. This is due to the private households providing information with respect to their debt obligations. In their framework, the "risk-free" assets include Cash, Deposit Account Balances, Canada Savings Bonds, and Personal Property. As in Cohn et al., the households were split into different wealth groups. What is interesting is that for poorer individuals they found evidence of IRRA while for the middle-wealth individuals DRRA was supported and lastly for the wealthiest group the findings pointed to CRRA. However, these findings are based on interpersonal comparisons of utility. A similar approach was done by Guiso et al. \cite{Guiso1996}. In fact, Guiso et al. revisited the RRA subject by doing a cross-sectional analysis on a random sample of 8,274 Italian households who provided information with respect to their portfolio allocation proportions. Their findings showcased that the households revealed DRRA.
	
	In 1994, Haim Levy \cite{Levy1994} introduced a different approach in determining the slope of RRA of individuals. Levy conducted an experiment in which 62 of his MBA students participated. Each student was given an initial investment allotment of $30,000\$$ "paper" money and was offered stocks of 20 pure equity firms as well as a "risk-free" rate of $2\%$. The major difference between Levy's approach and the two aforementioned works is that he ran time-series regressions for each individual and did not test cross-sectionally. What this means, is that through his approach the results are not based on interpersonal utility functions. Now, in terms of RRA, Levy argued that the results indicate a decreasing trend with respect to wealth. On top of that, Levy also did robustness checks studying the subjects' portfolio allocation when the only available assets are one risky and one risk-free asset. In this frame, the conclusions did not deviate from the previous findings. Even though this new approach overcomes some previous limitations an important caveat of this work is the small sample size.
	
	A whole different approach was proposed from Perignon and Villa \cite{Perignon2002} in 2002. More specifically, they attempted to extract information from Put and Call Options on CAC 40 with regards to the slope of RRA. The two authors, based their methodology on Ait-Sahalia and Lo (2000) \cite{Ait_Sahalia2000} who defined a pure-exchange economy. Namely, in this economy and in equilibrium, the investor optimally invests all his wealth in the single risky stock at every instant prior to the terminal date and then consumes the terminal value of the stock at time $T$. This implicitly says that the level of wealth of the individual is exactly equal to the stock price at each time. In other words, the representative agent consumes only at the final date and maximizes the expected utility of the terminal wealth by choosing the amount invested in the stock at each intermediary date. In this framework, Ait-Sahalia and Lo derive the "implied" RRA formula. Empirically, Perignon and Villa, obtain an "implied" RRA for the CAC 40 stock index and they conclude that as the index price (i.e. the wealth) increases the "implied" RRA decreases.
	
	\subsection{Increasing Relative Risk Aversion Literature}
	Contrary to the previous works, Siegel and Hoban \cite{Siegel1982} find that by restricting the sample to higher wealth target groups will lead to DRRA or CRRA. But they argue that "the use of a broader based sample and a more comprehensive measurement of wealth alters the conclusions and a pattern indicative of increasing relative risk aversion emerges". Siegel and Hoban used data from the US National Longitudinal Surveys (NLS). More specifically, they acquired a sample of 2,881 different sets of asset holdings of individual households. The proxy for level of wealth was Net Worth. As for the "riskless" assets they used Cash, Deposit Account Balances, and U.S. Savings Bonds. Splitting into different groups of level of wealth they concluded that for any group of individuals there is evidence of IRRA rather than DRRA. We should keep in mind that this paper constitutes a cross-sectional analysis as in \cite{Cohn1975} and \cite{Morin1983}.
	
	In 2003, Eisenhauer and Ventura \cite{Eisenhauer2003} found evidence of both DARA and IRRA. The authors asked different households in Italy the following question: "You are offered the opportunity of acquiring a security permitting you, with the same probabilities, either to gain 10 million lire or to lose all the capital invested. What is the most you are prepared to pay for this security?". The answer to this question is denoted by $z$ and we can interpret it as the CE of the respective household. The lower $z$ is the more risk-averse the household. The proxy of level wealth, denoted by $w$, was the average income of each household from 1993 to 1995. Now, the authors observed that 1,624 of the households interpreted the 10 million lire as a gross gain meaning that in the favourable case they get $w-z+10$ while in the other case their wealth becomes $w-z$. A number of 1,730 households interpreted the 10 million lire as a net gain meaning that in the favourable case they get $w+10$ while in the other case their wealth remains to be equal to $w$. In any case, they found that investors exhibit DARA and IRRA. Again, the most obvious limitations are that the results are based on a hypothetical survey question as well as on a cross-sectional analysis.
	
	Two additional works finding evidence of IRRA are those of \cite{Barsky1997} and \cite{Bar-Shira1997}. The first one was based on a set of gamble questions that had to do with the retirement decisions. The second one extracts the risk attitude of Israeli farmers.
	
	\subsection{Constant Relative Risk Aversion Literature}
	The research works finding evidence of CRRA are more recent and better substantiated. The first work finding evidence of CRRA was that of Friend and Blume (1975) \cite{Friend1975}. Using cross-section regressions based on household-level data on asset holdings the authors conclude that CRRA characterizes household behaviour. The main drawbacks of this study are that it is based on cross-sectional data and the data is only focused on high-wealth households.
	
	We will now discuss about the work of Sahm (2012) \cite{Sahm2012} which is the latest work of our review but it had started since 2006. Sahm initialized the use of panel data instead of cross-section data with regards to determining the slope of RRA in terms of wealth. In particular, he used gamble responses across the 1992 to 2002 waves of the Health and Retirement Study (HRS). The gamble under study is the following: "Suppose that you are the only income earner in the family. Your doctor recommends that you move because of allergies, and you have to choose between two possible jobs. The first would guarantee your current total family income for life. The second is possibly better paying, but the income is also less certain. There is a 50-50 chance the second job would double your total lifetime income and a 50-50 chance that it would cut it by a third. Which job would you take - the first job or the second job? Individuals who accept the first risky job then consider a job with a larger downside risk of one-half, while those who reject the first risky job are asked about a job with a smaller downside risk of one-fifth. If they reject the first two risky jobs, individuals consider a third risky job that could cut their lifetime income by one-tenth. Likewise, if they accept both risky jobs, individuals consider a third risky job that could cut their lifetime income by three-quarters." Sahm finds no effect of wealth changes on relative risk aversion and thus concludes CRRA.
	
	The next work we are going to discuss is that of Brunnermeier and Nagel (2008) \cite{Brunnermeier2008}. The authors use household-level panel data from the Panel Study of Income Dynamics (PSID), covering a period of 20 years between 1984 and 2003. The data includes Asset holdings, Income and households characteristics. To identify how wealth changes are related to market participation they use probit regressions. In particular, the authors first split the data  in two subperiods, 1984 to 1999 and 1999 to 2003. The reason is that for 1984–1999 the time-span between successive waves of the PSID with wealth information is $k=5$ years, and for the 1999–2003 sample the time-span between successive waves of the PSID with wealth information is $k=2$ years. In practice, the probit regressions derive the probability of the households who did not participate in the stock market at time $t-k$ to enter until time $t$. They also estimate the probability that a household that is participating at $t-k$ to exit the stock market until $t$. The authors find that for both subperiods, namely 1984-1999 and 1999-2003, there is a $1\%$ probability to enter the market when there is a $10\%$ increase in the wealth. Accordingly, the probability of exiting the stock market when there is an increase in the wealth is extremely low. Thus, they conclude that the households' relative risk aversion remains constant with respect to a change in wealth (CRRA). A major improvement of this work is that it uses panel data. This way you avoid the implicit assumptions made in the cross-sectional analysis. Namely, assuming that the distributions of wealth and preferences are independent.
	
	Another important work is that of Chiappori and Paiella (2011) \cite{Chiappori2011}. Similar to \cite{Brunnermeier2008} and \cite{Sahm2012}, they use panel data to showcase evidence of CRRA. In fact, they prove that previous studies that supported their findings on cross-sectional analysis are led to erroneous conclusions. Citing the authors: "without a priori restrictions on the joint distribution of wealth and preferences, the form of individual preferences simply cannot be recovered from cross-sectional data. In fact, any form of individual preferences is compatible with any observed, joint distribution of wealth and risky asset shares provided that one can freely choose the joint distribution of wealth and preferences.". In other words, estimating the joint distribution of wealth and risky asset shares using cross-section data derives the joint distribution of wealth and preferences only and only if you pre-assume that preferences are the same for each investor. This of course is an over-simplification of reality, which substantiates the use of panel data instead of cross-section data. The data used is from the Survey of Household Income and Wealth (SHIW), which is a large-scale household survey run every two years by the Bank of Italy from which they get asset allocations for 1,332 households. Also, they exclude all the households with a change in wealth less than $25\%$. As in \cite{Brunnermeier2008} and \cite{Sahm2012}, they find evidence of CRRA.
	
	\section{Degree Of Risk Aversion For A Lottery With Non-Zero Mean}
	In the previous section, we derived the "absolute" and "relative" risk premium for a "fair" lottery. In reality however, stock prices are not a fair lottery meaning that $E[Z]=\mu_Z\neq 0$. Equivalently, stock returns are not a "fair" lottery and so $E[R]=\mu_R\neq 0$. Deriving ARA under $\mu_Z\neq 0$ leads to RRA through $\lambda(w_0)=w_0r(w_0)$. So, for a non-fair lottery with $\mu_Z\neq 0$ we have
	\begin{equation*}
		\rho(W_1)=\rho(w_0+Z)=E[w_0+Z]-z_0=w_0+\mu_Z-z_0,
	\end{equation*}
	where $z_0$ represents the CE of lottery $Z$ with $w_0$ level of wealth (or equivalently the CE of lottery $w_0+Z$). Thus, the CE will be
	\begin{equation*}
		z_0=w_0+\mu_Z-\rho(w_0+Z).
	\end{equation*}
	According to Definition (3.1), the following equation must be satisfied
	\begin{equation*}
		E[U(Z)]=U(z_0).
	\end{equation*}
	In our case,
	\begin{equation*}
		E[U(w_0+Z)]=U(w_0+\mu_Z-\rho(w_0+Z)).
	\end{equation*}
	\noindent
	At this point, we use Taylor expansions of 2nd-order around $w_0$ for both sides of the equation. For the left side of the equation we derive
	\begin{align}
		\begin{split}
			E[U(w_0+Z)] &\approx U(w_0)+U'(w_0)E[w_0+Z-w_0]+\frac{U''(w_0)}{2}E[(w_0+Z-w_0)^2]\\
			&\approx U(w_0)+U'(w_0)\mu_Z+\frac{U''(w_0)}{2}(\sigma_Z^2+\mu_Z^2).
		\end{split}
	\end{align}   
	\noindent
	Accordingly, for the right side of the equation we get
	{\footnotesize
		\begin{align}
			\begin{split}
				U(w_0+\mu_Z-\rho(w_0+Z)) &\approx U(w_0)+U'(w_0)(w_0+\mu_Z-\rho(w_0+Z)-w_0)+\frac{U''(w_0)}{2}(w_0+\mu_Z-\rho(w_0+Z)-w_0)^2\\
				&\approx U(w_0)+U'(w_0)(\mu_Z-\rho(w_0+Z))+\frac{U''(w_0)}{2}(\mu_Z-\rho(w_0+Z))^2\\
				&\overset{\text{drop 3rd term}}{\approx} U(w_0)+U'(w_0)(\mu_Z-\rho(w_0+Z)).
			\end{split}
		\end{align}
	}
	\noindent
	Now, combining (3) and (4) we get
	\begin{align*}	
		&U(w_0)+U'(w_0)\mu_Z+\frac{U''(w_0)}{2}(\sigma_Z^2+\mu_Z^2) \approx U(w_0)+U'(w_0)(\mu_Z-\rho(w_0+Z))\\
		&\frac{U''(w_0)}{2}(\sigma_Z^2+\mu_Z^2) \approx -U'(w_0)\rho(w_0+Z)\\
		&\rho(w_0+Z)\approx -\frac{U''(w_0)}{2U'(w_0)}(\sigma_Z^2+\mu_Z^2)\\
		&\rho(w_0+Z)\approx \frac{1}{2}r(w_0)(\sigma_Z^2+\mu_Z^2).
	\end{align*}
	So, compared to (1) the risk premium will satisfy the following equation
	\begin{equation}
		\boxed{\rho(w_0+Z)\approx \frac{1}{2}r(w_0)(\mu_Z^2+\sigma_Z^2)} 
	\end{equation}
	The main change is that the risk-premium is now positively related also to $\mu_Z^2$. In other words, a higher mean in absolute terms, would increase the required $\rho$. Now solving for the ARA term $r(w_0)$ we get
	\begin{equation*}
		w_0+\mu_Z-z_0\approx \frac{1}{2}r(w_0)\mu_Z^2+\frac{1}{2}r(w_0)\sigma_Z^2
	\end{equation*}
	\begin{equation}
		\boxed{r(w_0)\approx \frac{2(w_0+\mu_Z-z_0)}{\mu_Z^2+\sigma_Z^2}} 
	\end{equation}
	The above formula represents a closed-form solution for ARA when $Z$ is not a "fair" lottery. In relative terms, $Z$ can be expressed in the following way
	\begin{equation*}
		R=\frac{Z}{w_0}\Leftrightarrow Z=Rw_0.
	\end{equation*}
	Therefore, we have 
	\begin{align*}	
		&\rho(w_0+Z)\approx \frac{1}{2}r(w_0)w_0^2(\sigma_R^2+\mu_R^2)\\
		&\rho(w_0+Z)\approx \frac{1}{2}r(w_0)(\sigma_Z^2+\mu_Z^2).
	\end{align*}
	And so, the "relative" risk premium for $\mu_R\neq 0$ takes the following form
	\begin{align*}	
		\frac{\rho(w_0,Rw_0)}{w_0}=\tilde{\rho}(w_0,R)&\approx \frac{1}{2}r(w_0)w_0(\sigma_R^2+\mu_R^2)\\
		&\approx \frac{1}{2}\lambda(w_0)(\sigma_Z^2+\mu_Z^2)
	\end{align*}
	\begin{equation}
		\boxed{\tilde{\rho}(w_0,R)=\frac{1}{2}r(w_0)w_0(\mu_R^2+\sigma_R^2)} 
	\end{equation}
	Now, we can derive the RRA term $\lambda(w_0)$.
	\begin{equation}
		\boxed{\boldsymbol{\lambda(w_0)=\frac{2(1+\mu_R-\tilde{z}_{Rw_0})}{\mu_R^2+\sigma_R^2}}} 
	\end{equation}
	We are going to use this last formula to extract the level of RRA of the market. Observe that $\mu_R$ and $\sigma_R$ could represent  the mean and standard deviation of market returns, respectively. Likewise, $\tilde{z}_{Rw_0}$ represents the risk-free rate. Finally, $w_0$ is the market capitalization of the market. But before going any further, a careful overview of the existing literature needs to be made.
	
	\section{Empirical Results}
	The previous sections showcased that there is no definitive answer as to if and how much RRA is affected by a change in wealth. In our view, the largest part of the literature has focused on determining the slope of RRA using an approach with multiple perils. Namely, most studies have to deal with demographic and socio-economic characteristics or over-simplifications. Even controlling for those characteristics may lead to sub-optimal conclusions. The most recent part of the literature finds evidence of CRRA using panel data. We believe that these studies are more meaningful and better substantiated. 
	
	In our framework, we try to avoid the issues noted in cross-sectional data and at the same time simplify even further the extraction of RRA from real data. More specifically, we believe that one should focus on determining the market's ARA and RRA through the formulae we derived. These formulae are free of any need of including and subsequently controlling for any subjective characteristic of the investors such as educational status or age or even level of wealth. The only subjective characteristic lies in the CE (or equivalently, risk-free rate) which we will measure using 10-year Treasury yields, since this is considered a logical proxy of the risk-free rate in the literature.
	
	We use monthly returns and market cap from four different markets, namely CAC 40, EURO, S\&P 500 and STOXX 600, spanning from 1991 to 2021 excluding 2008. The reason why we exclude 2008 is because this period is known for extreme levels of volatility which may affect our concluding remarks. We believe that the advantage of using monthly returns instead of daily returns is that they do not exhibit volatility clustering. This means that $\mu_R$ and $\sigma_R$ can be estimated recursively. Estimations start from 2012 using all the past data and continue up until 2022. As a proxy of risk-free rate we use the respective 10-year Treasury yields of each market. The reason we want to extract RRA for all these indexes is to determine whether or not the results are consistent across different markets.
	
	The first thing we want to answer is how do ARA and RRA of the aforementioned market indices change with respect to an increase in their market cap. From Figure 5, we deduce that for all market indices ARA is decreasing (DARA). In fact, the correlation is almost perfectly negative for all markets. What this says, is that in all these indices the investors become more risk averse in absolute terms. So, as the level of wealth (market cap) increases investors are willing to take on more risk. This evidence is strongly supported by the findings of the literature. Figure 6, presents the RRA of each market with respect to each level of wealth. We observe that for the European indices RRA displays an increasing trend with high positive correlation. When it comes to S\&P 500, we see that the correlation is $-36\%$, but we should not focus entirely on it. Taking a closer look, we can see that up until the level of a market cap of approximately $27$ trillion dollars the RRA exhibits an increasing trend. More specifically, the correlation between the market cap and the degree of RRA up until this point is approximately $90\%$. In fact, we argue that the sudden drop at a market cap of approximately $27$ trillion dollars is due to an important economic event that took place during the first quarter of 2020, which we will discuss, in detail, in the following paragraph. Thus, we maintain that we should not be distracted by a one-off event.
	
	A closer look at Figure 1 justifies the sharp drop observed in the level of relative risk aversion of the US market. In this figure, we plot RRA and Market Cap in time. In February 2020, the Fed Chair Mr. Jerome H. Powell, in an attempt to soothe the market which was extremely nervous due to the COVID-19 pandemic, hinted a forthcoming rate cut. In March 2020, Fed announced the new QE program through which it would purchase \$600 billion in bank debt, U.S. Treasury notes, and mortgage-backed securities (MBS) from member banks. At the same time, it cut down the federal funds rate by a total of 1.5 percentage points. To comprehend the impact of such an action by the central bank of United States we should refer to equation (7). The level of RRA is negatively related to the CE (or equivalently risk-free rate) which is set by the investors. So, assuming that the objective (probabilistic) characteristics of the lottery remain unchanged, as CE gets smaller RRA should increase. Equivalently, the investors would become more risk-averse and so they would require a smaller CE in order to avoid the risky lottery. But this is not the case in the graph. In fact, RRA plummets. To determine the reason behind this drop we should break down the graph in two parts. Up until February 2020, it was the market that had been setting the CE. But in March 2020, Fed set a new lower level of risk-free rate. This new rate was not the result of a change in the investors' level of risk-aversion. It actually resulted from the artificial cut in interest rates by the Federal Reserve Bank. This led to a group of less risk-averse bondholders shifting to the stock market. So, the drop in RRA gives the impression that all bondholders became less risk-averse, which is not true. Assuming that Fed had not intervened, in which case would we have an analogous shift from bonds to stocks by the bondholders? In case the bondholders became more risk-averse. Thus, we conclude that the QE program forced investors to move into risky assets and acted "as-if" the bondholders became more risk-averse. Furthermore, this action seems that affected the other markets as well, which proves that markets are interconnected when it comes to important economic events.
	
	The discussion made in the previous paragraphs leads to the following conclusion. The investors reveal DARA and IRRA for 2012-2022. It needs to be highlighted that this conclusion applies for this time period and should not be treated as a generalization on the investors' risk preferences. In other words, we imply that market participants may have had different risk preferences during different time periods. In fact, we will see later on that for different time periods the slope of the RRA of the different market indices varies. These findings indicate that, currently, we could assume that the investors' class of utility functions should satisfy both DARA and IRRA. The following utility functions are a part of this class.
	\begin{align*}
		U(W)=(W+a)^c, \ a>0, \ &0<c<1, \quad U(W)=-(W+a)^{-c}, \ a>0, \ c>0,\\
		&U(W)=log(W+a), \ a>0
	\end{align*} 
	Any utility function from this group serves as an appropriate proxy for the true utility function of the markets. We proceed as follows. First, we will compare our evidence of DARA and IRRA to a rolling-window approach. This will serve as a robustness test of our evidence based on recursive estimations. Then, we will extract the investors' risk preferences through S\&P 500 for different time periods, namely, 1993-1998 and 1999-2005.
	
	\begin{figure}[H]
		\includegraphics[scale=0.8]{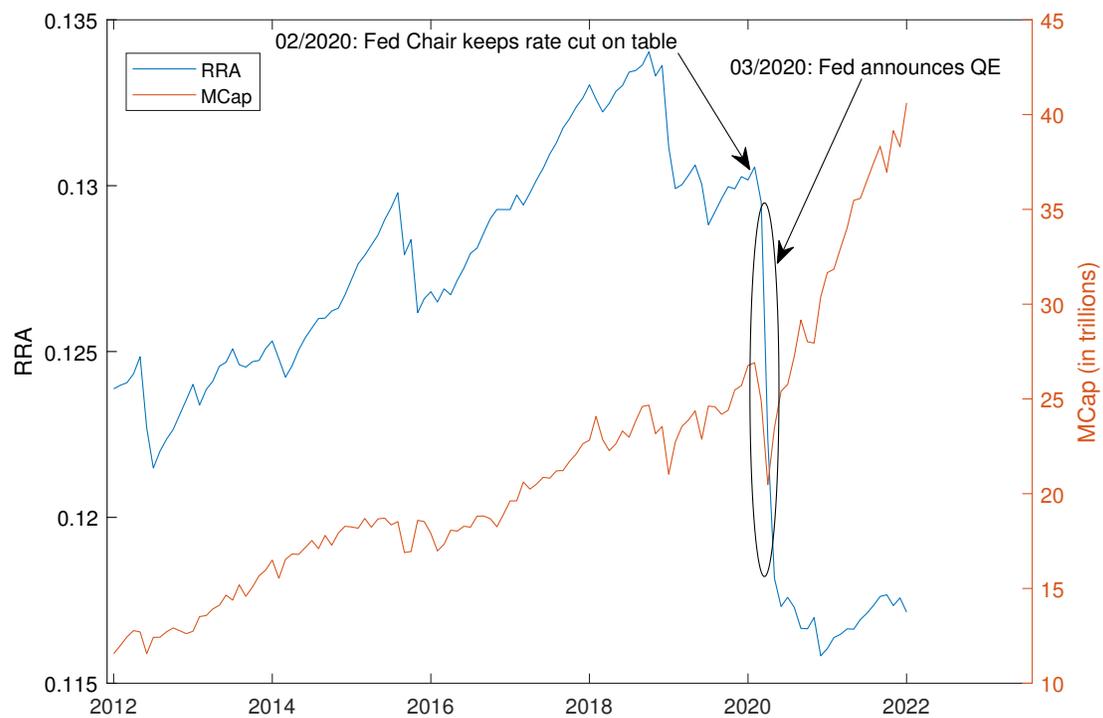}
		\caption{RRA: S\&P 500}
	\end{figure}
	
	\newpage
	\subsection{Rolling-Window Approach}
	As previously stated, our estimations with regards to $\mu_R$ and $\sigma_R$ were done recursively. One could argue that such an approach is problematic, reasoning that the estimations should better be updated using more recent data. For that reason, we also estimate $\mu_R$ and $\sigma_R$ by employing a rolling-window approach. We test three different window sizes, $M=60$ (5-years), $M=120$ (10-years) and $M=180$ months (15-years). These three window sizes are regularly applied by the literature for monthly data. 
	
	The graphs in Figures 8 and 9 showcase that for a window size of $M=60$ months, or equivalently, a $5$-year window, there is strong evidence of DARA and IRRA. In fact, the correlation between the ARA and the market cap is highly negative and smaller than $-30\%$ for all markets. Accordingly, the correlation between the RRA and the market cap is above $40\%$ for all markets except for S\&P 500, for which the magnitude of correlation is mainly affected by the QE program announced by Fed. When we increase the window size to $M=120$ motnhs, or equivalently, a $10$-year window, Figures 10 and 11, point to the same direction with the exception of S\&P 500, in which case it seems that the RRA has a decreasing trend. Lastly, in Figures 12 and 13, it is evident that we have similar conclusions to those made in the 5-years case. More specifically, for a window size of $M=180$ months, or equivalently, a $15$-year window, the evidence of DARA and IRRA is even stronger, with correlations being higher in absolute terms. Overall, our results for 2012-2022 reveal strong evidence of DARA and IRRA.
	
	\subsection{RRA And ARA For S\&P 500 In Different Time Periods}
	Now, we are going to study ARA and RRA of S\&P 500 in different time periods. Doing that will help us determining the risk attitude of US investors across the years. On top of that, we would like to check whether or not the evidence of DRRA and CRRA in earlier studies is supported by our model. For this, we will recover ARA and RRA for 1993-1998 and 1999-2005. 
	
	Figure 2 displays the level of ARA and RRA for 1993-1998. What is worth noticing is that investors' risk attitude was rather different than currently is, revealing DRRA instead of IRRA. This result supports the evidence of DRRA found in \cite{Morin1983}, \cite{Levy1994}, \cite{Guiso1996} and \cite{Perignon2002} for earlier time periods. Although some of these works refer to different time periods or different markets they are indicative of the investors' risk attitude overall. Now, the RRA graph on Figure 3 is harder to interpret. The graph is indecisive between CRRA and IRRA. Studies like those of \cite{Sahm2012}, \cite{Brunnermeier2008} and \cite{Chiappori2011} found evidence of CRRA for this time period. However, our graph does not offer a clear conclusion. In terms of the ARA, both Figures reveal DARA which has been validated by most empirical studies, for any time period. 
	
	Overall, our empirical evidence demonstrates the importance of studying the risk attitude of investors across different periods. It seems that the market's utility function varies through years. More specifically, our findings point to a utility function satisfying DRRA and DARA for years 1993 through 1998 meaning that investors were willing to take on more risk as they became wealthier. For 1999 to 2005, investors started being more cautious as they reached higher levels of wealth, revealing CRRA/IRRA and DARA. Currently, the different markets seem to become continuously more risk-averse as their market caps grow even further, revealing evidence of IRRA and DARA.
	
	\begin{figure}[htp]
		\begin{subfigure}{0.45\textwidth}
			\includegraphics[width=\linewidth]{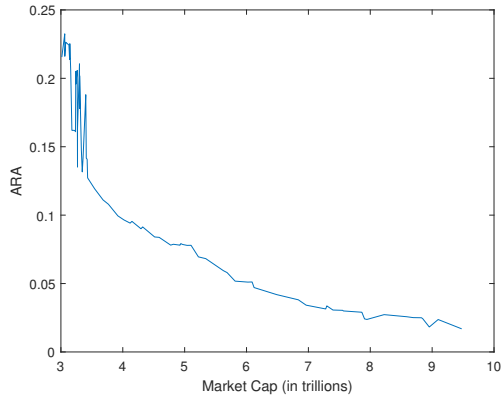}
			\caption{ARA}
		\end{subfigure}
		\hspace*{1in}
		\begin{subfigure}{0.45\textwidth}
			\includegraphics[width=\linewidth]{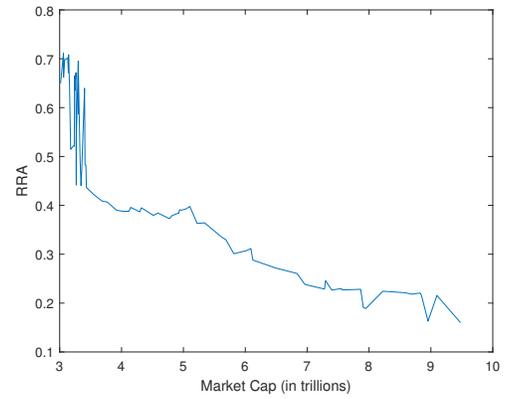}
			\caption{RRA}
		\end{subfigure}
		\caption{ARA and RRA: 1993-1998}
	\end{figure}
	
	\begin{figure}[htp]
		\begin{subfigure}{0.45\textwidth}
			\includegraphics[width=\linewidth]{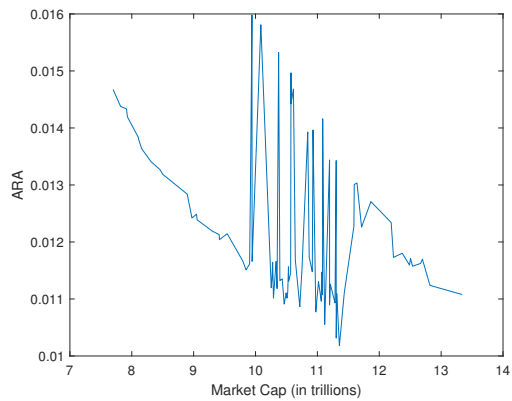}
			\caption{ARA}
		\end{subfigure}
		\hspace*{1in}
		\begin{subfigure}{0.45\textwidth}
			\includegraphics[width=\linewidth]{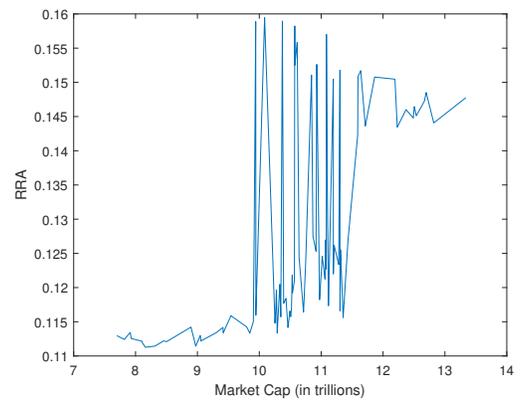}
			\caption{RRA}
		\end{subfigure}
		\caption{ARA and RRA: 1999-2005}
	\end{figure}
	\newpage
	
	Following our empirical evidence in the previous sections, a subsequent question would be: How much does a wrong assumption with respect to the utility function of an investor, made by an asset manager, will affect the structure of his portfolio? Could we find an easy and intuitive way to measure  the difference between the level of risk that the investor should incur, based on his "true" utility function, and the level of risk that the asset manager recommends, based on his perception of the investor's utility function? We approach these questions in the following section.
	
	\subsection{Measuring The Differences In Portfolio Diversification Among Different Utility Functions}
	Consider the following case. A financial advisor designs a survey to extract the risk preferences of an investor. The asset manager misinterprets the investor's answers and concludes that the investor reveals IARA and IRRA. In reality however, the investor reveals DARA and IRRA. How much would the decisions made by the asset manager on behalf of the investor be affected? Assuming IARA and IRRA leads to a very well-known utility function, the quadratic $U(W)=W-bW^2$ with $b>0$. This utility function is considered by the literature to be a sufficient condition for Markowitz's Mean-Variance Optimization method. So, by assuming a quadratic utility the asset manager should use Markowitz's method to diversify the investor's portfolio. In the following paragraph we are going to illustrate an intuitive way to measure the effect of assuming a quadratic rather than a DARA and IRRA utility function like one of those shown below.
	\begin{align*}
		&U(W)=(W+a)^c, \ a>0,  \ 0<c<1, & U(W)=-(W+a)^{-c}, \ a>0, \ c>0,\\
		&U(W)=log(W+a), \ a>0 \ & U(W)=-e^{-c(W+a)} , \ a>0, \ c>0
	\end{align*}
	
	Assume that at time $t=0$ the investor's wealth is $w_0$. The asset manager decides to invest $w_s$ in a risky asset with return $R$ and $w_0-w_s$ in a riskless asset with rate of return $R_f$. Both $w_0$ and $w_s$ are measured in percentages. So, $w_0=100\%$ while $w_s$ can take any value. The idea is that if $w_s>w_0$, the asset manager can borrow money to over-invest in the risky asset, while if $w_s<0$ the asset manager will short-sell the risky asset. So, at point $t=1$ the investor's wealth will be
	\begin{align*}
		W_1&=w_s(1+R)+(w_0-w_s)(1+R_f)\\
		&=w_0(1+R_f)+w_s(R-R_f).
	\end{align*}
	Based on VN-M Representation Theorem the asset manager will simply need to maximize the expected utility function of the investor for $W_1$.
	\begin{equation*}
		\underset{W_1}{\max}E[U(W_1)]\Leftrightarrow E[U'(W_1)]=0
	\end{equation*}
	Solving the above equation derives $w_s$. This percentage will differ among different utility functions which means that we can measure the differences between them. In our case, we will find $w_s^{quad}$ from the quadratic utility function $U(W)=W-bW^2$ with $b=0.2,0.3,0.4$ and compare it with $w_s^{log}$ from $U(W)=\log{W}$. Below, we derive the two quantities. Namely, for $U(W)=W-bW^2$ we have
	\begin{equation*}
		w^{quad}_s = \frac{\mu_R-R_f-2b(1+R_f)(\mu_R-R_f)}{2b(\mu_R^2+\sigma_R^2-2\mu_RR_f+R_f^2)}
	\end{equation*}
	In terms of $U(W)=\log{(W)}$, we are unable to apply the approach we propose directly on it, since we will end up with a fraction equal to $0$ that cannot be solved. We can omit this hurdle by taking the Taylor expansion of 2nd-order around $0$ as follows.
	\begin{align*}
		U(w_s)&\simeq U(0) + w_sU'(0) + \frac{w_s^2}{2}U''(0)\\
		&\simeq \log{(1+R_f)} + w_s\frac{R-R_f}{1+R_f} - w_s^2\frac{(R-R_f)^2}{2(1+R_f)^2} 
	\end{align*}
	Thus, we obtain
	\begin{equation*}
		w^{log}_s = \frac{(1+R_f)(\mu_R-R_f)}{\mu_R^2+\sigma_R^2-2\mu_RR_f+R_f^2}
	\end{equation*}
	To measure the difference between the two utility functions we need to use real data. For that reason, we are going to use monthly returns from S\&P 500 and 10-year Treasury yields spanning from 2012 to 2022. Then, we will plot only the weight invested in the risky asset, namely $w_s$, for both the quadratic and the logarithmic utility functions, with respect to time.
	\begin{figure}[H]	
		\includegraphics[width=\linewidth]{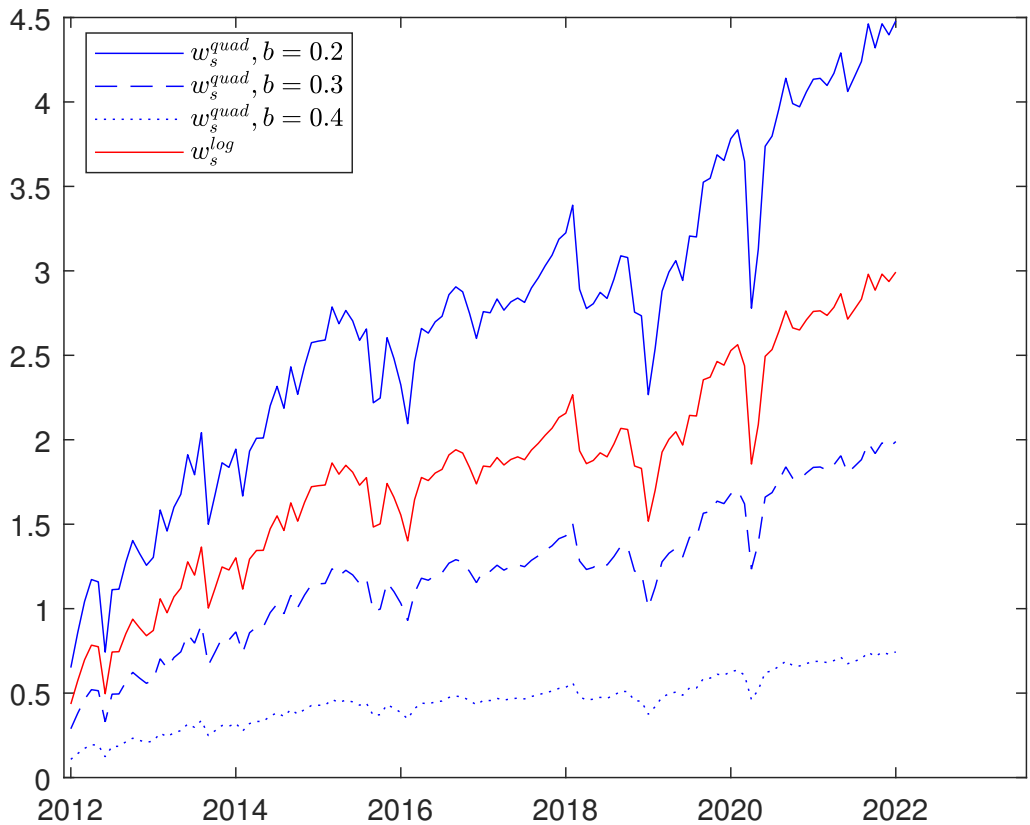}
		\caption{Quadratic vs Logarithmic utility's weight in the risky asset}
	\end{figure}
	The graph indicates that in case the asset manager had assumed one of the depicted parametrizations of the quadratic utility function for the investor, he would advise him to put significantly different portions of his wealth in S\&P 500 each month. Specifically, if $b=0.2$ the asset manager would underestimate the "true" level of risk-aversion of the investor, by advising him to put $50\%$ more on S\&P 500 than what his "true" level of risk tolerance suggests. While if $b=0.3,0.4$, the asset manager would overestimate the "true" level of risk-aversion of the investor, by almost $50\%$ to $100\%$. So, the asset manager's misjudgment would lead the investor to incur substantially more or less risk than he could actually tolerate. We should highlight here that even if we consider the case in which the "true" utility function of the investor is quadratic with $b=0.3$, the weight invested in the risky asset, namely $w_s$, is substantially higher through all years, compared to the $b=0.4$ case. So, even if the asset manager correctly detects that the investor's utility is quadratic, there is still plenty of room to underestimate (or overestimate) his true level of risk-aversion. 
	
	In the previous analysis, we compared the logarithmic utility function which is known to correspond to investors exhibiting DARA and IRRA, with the quadratic utility function which exhibits IARA and IRRA. So, although the logarithmic and the quadratic utility functions are both increasing and concave, they differ in the slope of the level of ARA. It would be interesting to compare utility functions that exhibit DARA and IRRA. So, instead of the quadratic utility function, we will now compare $\log{W}$ to $\sqrt{W}$ and $-e^{-2W}$.
	
	As in the case of $\log{W}$, we will need to use the Taylor expansion of 2nd-order around $0$ for $\sqrt{W}$, as follows.
	\begin{align*}
		U(w_s)&\simeq U(0) + w_sU'(0) + \frac{w_s^2}{2}U''(0)\\
		&\simeq \sqrt{1+R_f} + w_s\frac{R-R_f}{2\sqrt{1+R_f}} - w_s^2\frac{(R-R_f)^2}{8(1+R_f)^{3/2}} 
	\end{align*}
	Accordingly for $-e^{-2W}$, the Taylor expansion of 2nd-order around $0$ will be
	\begin{align*}
		U(w_s)&\simeq U(0) + w_sU'(0) + \frac{w_s^2}{2}U''(0)\\
		&\simeq -e^{-2(1+R_f)} + 2w_s(R-R_f)e^{-2(1+R_f)} - 2w_s^2(R-R_f)^2e^{-2(1+R_f)} 
	\end{align*}
	Thus, we derive
	\begin{align*}
		&w^{sqrt}_s = \frac{2(1+R_f)(\mu_R-R_f)}{\mu_R^2+\sigma_R^2-2\mu_RR_f+R_f^2}=2w^{log}_s\\
		&w^{exp}_s = \frac{\mu_R-R_f}{2(\mu_R^2+\sigma_R^2-2\mu_RR_f+R_f^2)}=\frac{w^{log}_s}{2(1+R_f)}
	\end{align*}
	The above formulae measure exactly the differences between the weight invested in the risky asset for each utility function. Namely, the square root utility function requires double the weight invested in the risky asset, compared to that of the logarithmic. Accordingly, the exponential utility function which represents a more risk-averse investor requires approximately half the weight invested in the risky asset, compared to that of the logarithmic. Below we plot the weight invested in the risky asset, namely $w_s$, for the different utility functions, with respect to time.
	\begin{figure}[H]	
		\includegraphics[width=\linewidth]{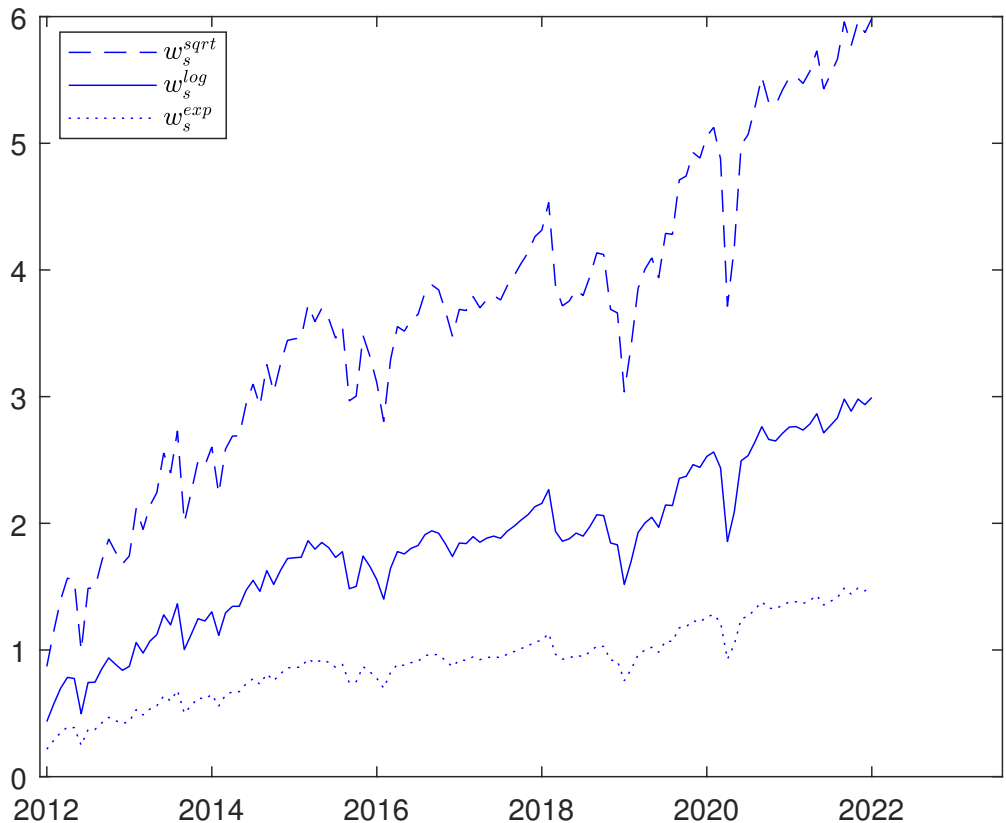}
		\caption{Logarithmic vs Square root vs Exponential utility's weight in the risky asset}
	\end{figure}

	To conclude, we designed a simple and intuitive method to measure the differences between alternative utility functions, with relative accuracy. A natural extension of this idea would be to study the above approach for DARA and IRRA utility functions, by using higher orders of Taylor series.
	
	\section{Conclusions}
	
	In this work, we extensively overviewed the various ways the literature has extracted the ARA and the RRA of investors through the years. We saw that, the literature universally accepts DARA but is not conclusive with respect to the RRA of investors. In fact, one part of previous research works reveals evidence of DRRA, another part finds evidence of IRRA and the most recent works point to CRRA. We asserted that using cross-section data as was regularly done in earlier research works, conceals risks. More specifically, cross-sectional analysis implicitly says that all investors have the same utility function, which is a rather unrealistic assumption. Studies basing their evidence on either panel data or Options markets data are better supported.
	
	We proposed a different approach in extracting both ARA and RRA. This approach, is based on carefully analyzing the Theory of Arrow-Pratt together with the different notions that underlie the Decision Theory under Risk. In particular, we derived closed-form expressions for the ARA and the RRA of investors, for non-"fair" lotteries. We proceeded with collecting data from different markets which we then applied on our formulae. Our findings, pointed out that through 2012-2022, European as well as US markets revealed evidence of DARA and IRRA. We further showcased that our formula captures important economic events as in the case of the QE program announced by Fed in March 2020, which caused a sharp drop in the investors' level of relative risk-aversion. Our results were further tested using a rolling-window approach for different window sizes $M=60$, $M=120$ and $M=180$ months. The rolling-window results supported our evidence of DARA and IRRA for 2012-2022. Then, we found out that for different time periods, namely, 1993-1998 and 1999-2005 RRA may vary. More specifically, for 1993-2005 we found strong evidence of DARA and DRRA while for 1999-2005 we found strong evidence of DARA but in terms of RRA our results were indecisive between CRRA and IRRA. In the last part of our work, we proposed a simple way to measure the effect of a wrong assumption with respect to the utility function of an investor. As it became clear, an investor with a logarithmic utility function is led to a very different portfolio structure compared to an investor with a quadratic utility function.
	
	\newpage
	\section*{Appendix}
	\begin{figure}[htp]
		\begin{subfigure}{0.45\textwidth}	
			\includegraphics[width=\linewidth]{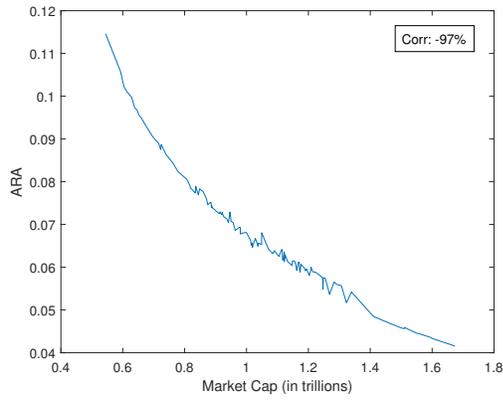}
			\caption{CAC 40}
		\end{subfigure}
		\hspace*{1in}
		\begin{subfigure}{0.45\textwidth}
			\includegraphics[width=\linewidth]{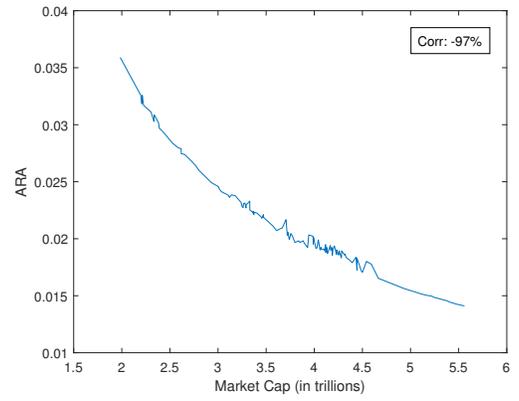}
			\caption{EURO}
		\end{subfigure}
		\begin{subfigure}{0.45\textwidth}
			\includegraphics[width=\linewidth]{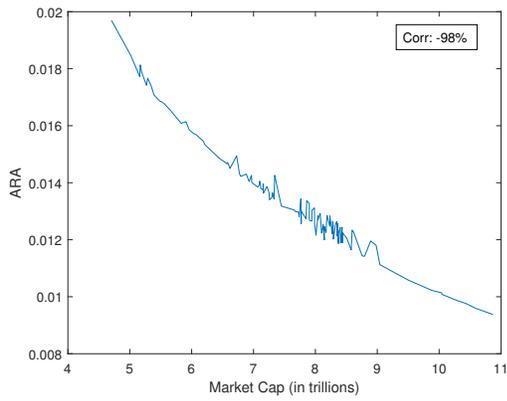}
			\caption{STOXX 600}
		\end{subfigure}
		\hspace*{1in}
		\begin{subfigure}{0.45\textwidth}
			\includegraphics[width=\linewidth]{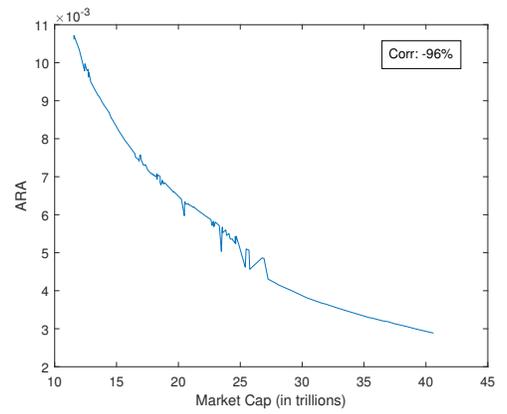}
			\caption{S\&P 500}
		\end{subfigure}
		\caption{ARA with sorted wealth}
	\end{figure}
	
	\newpage
	\begin{figure}[htp]
		\begin{subfigure}{0.45\textwidth}	
			\includegraphics[width=\linewidth]{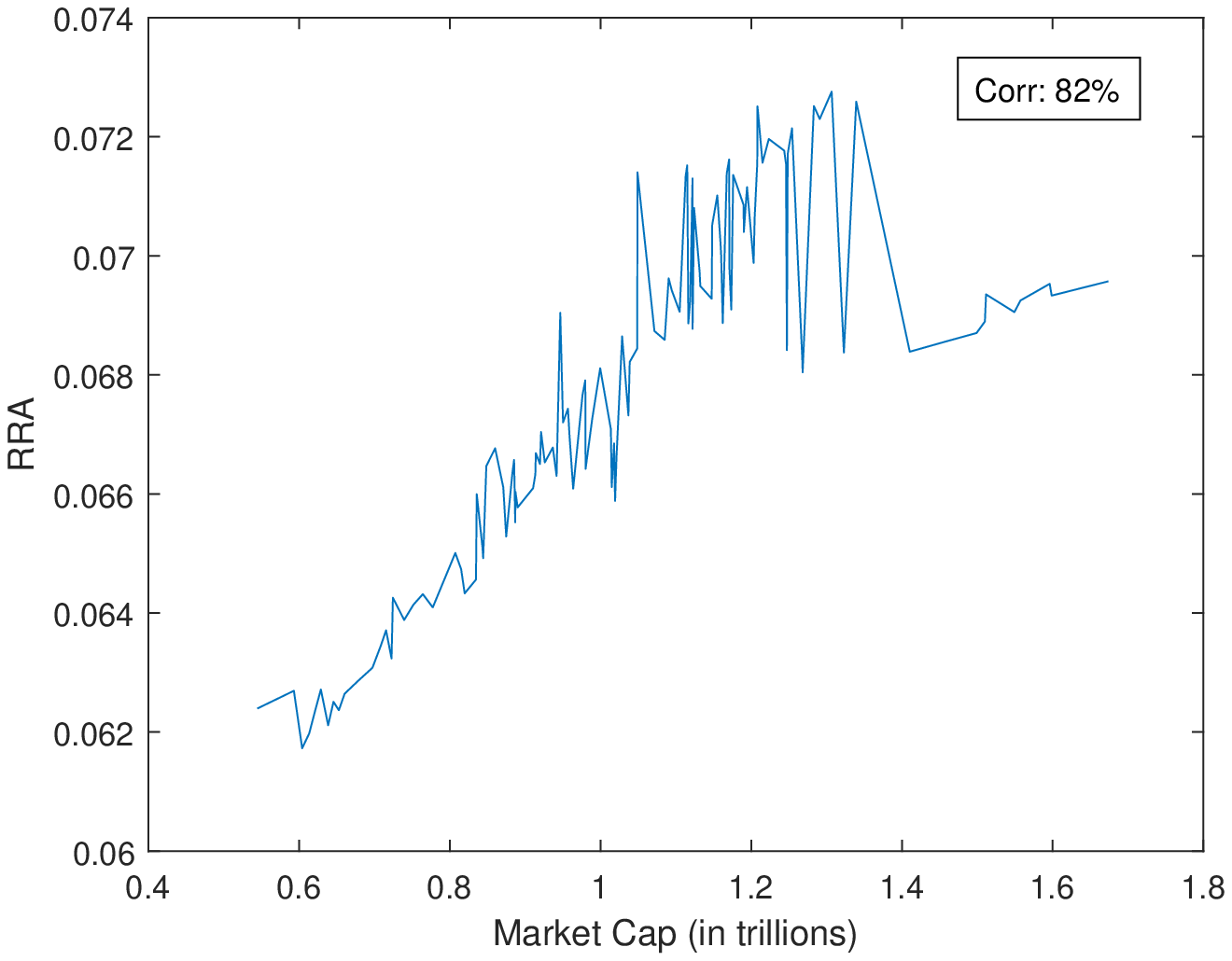}
			\caption{CAC 40}
		\end{subfigure}
		\hspace*{1in}
		\begin{subfigure}{0.45\textwidth}
			\includegraphics[width=\linewidth]{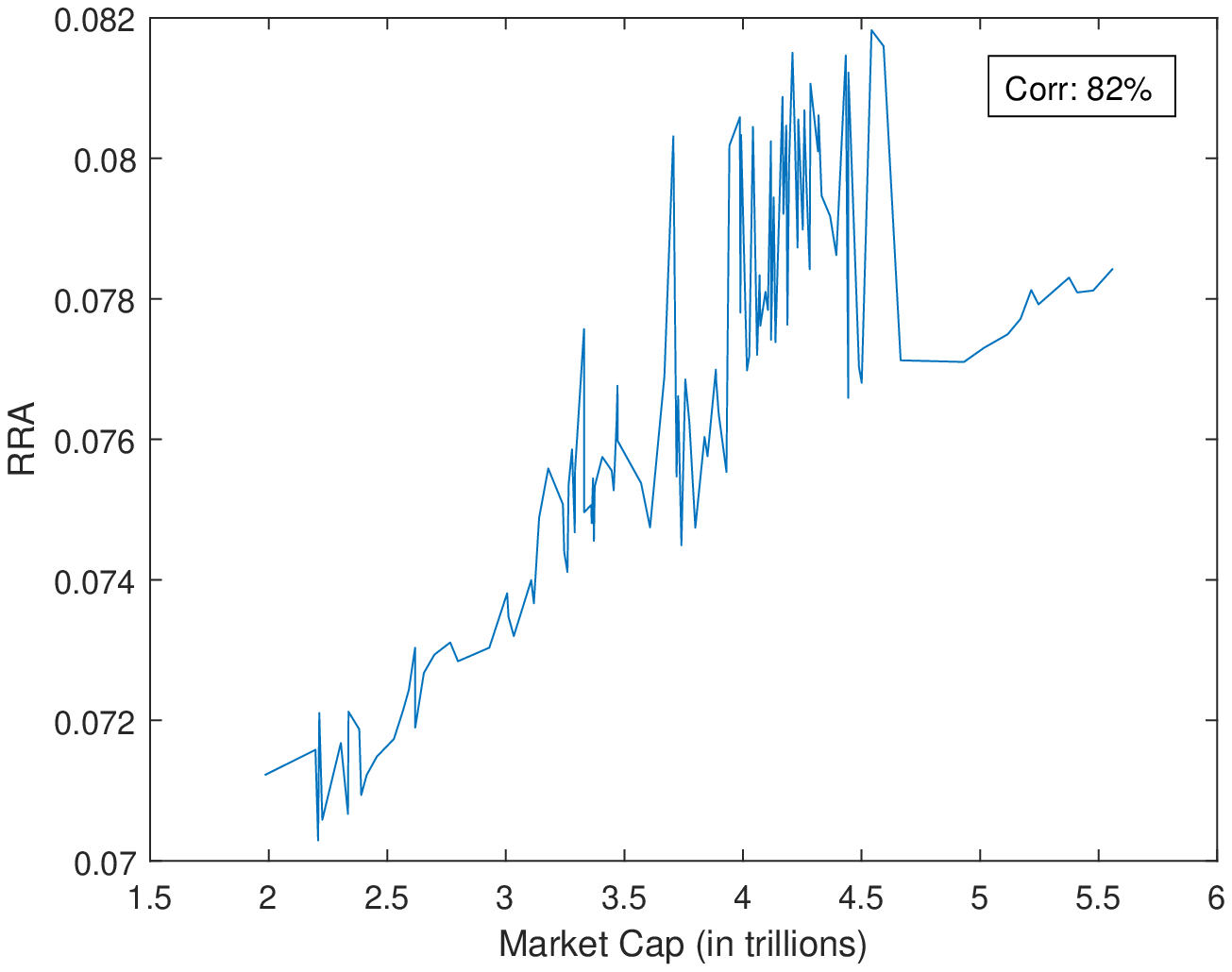}
			\caption{EURO}
		\end{subfigure}
		\begin{subfigure}{0.45\textwidth}
			\includegraphics[width=\linewidth]{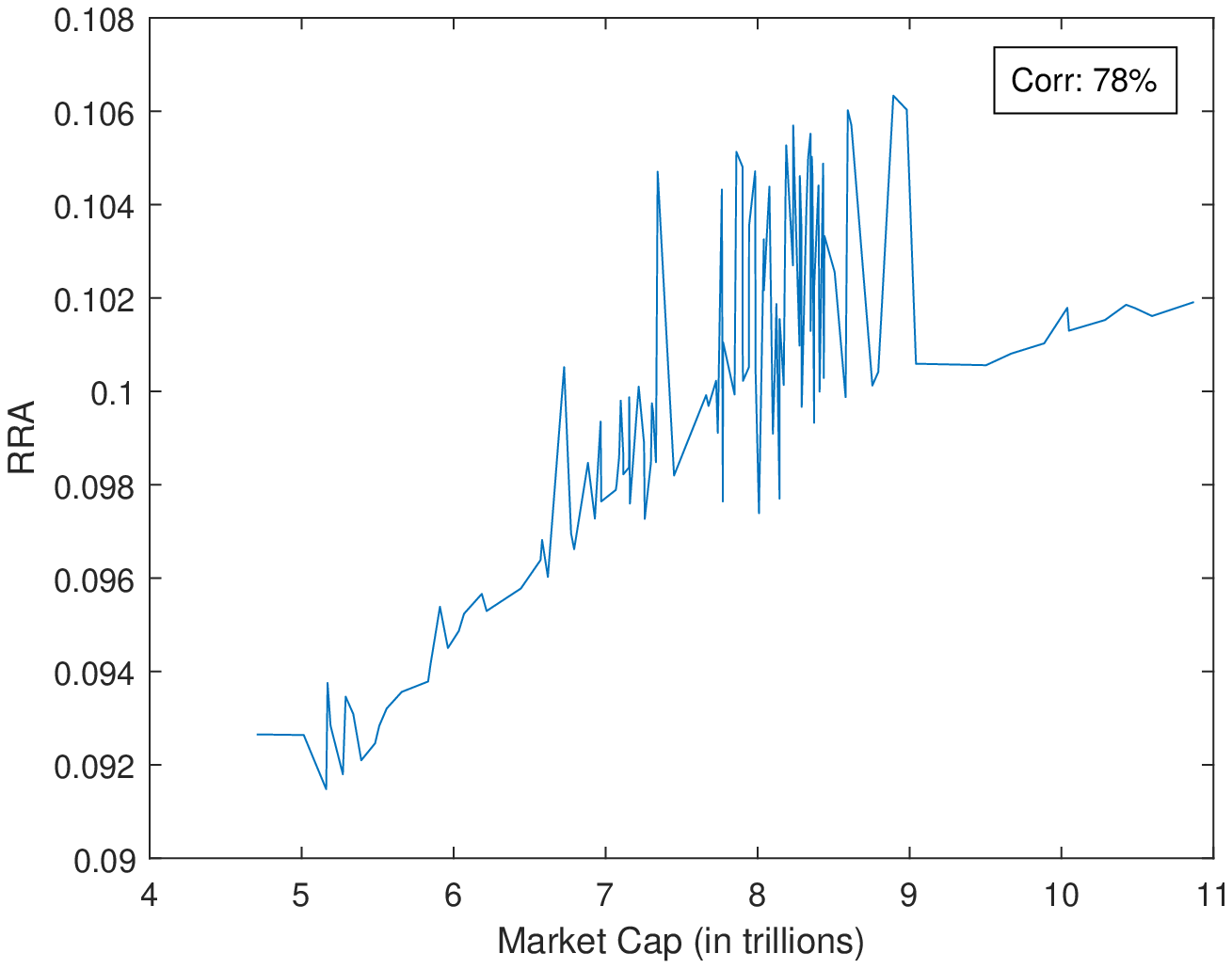}
			\caption{STOXX 600}
		\end{subfigure}
		\hspace*{1in}
		\begin{subfigure}{0.45\textwidth}
			\includegraphics[width=\linewidth]{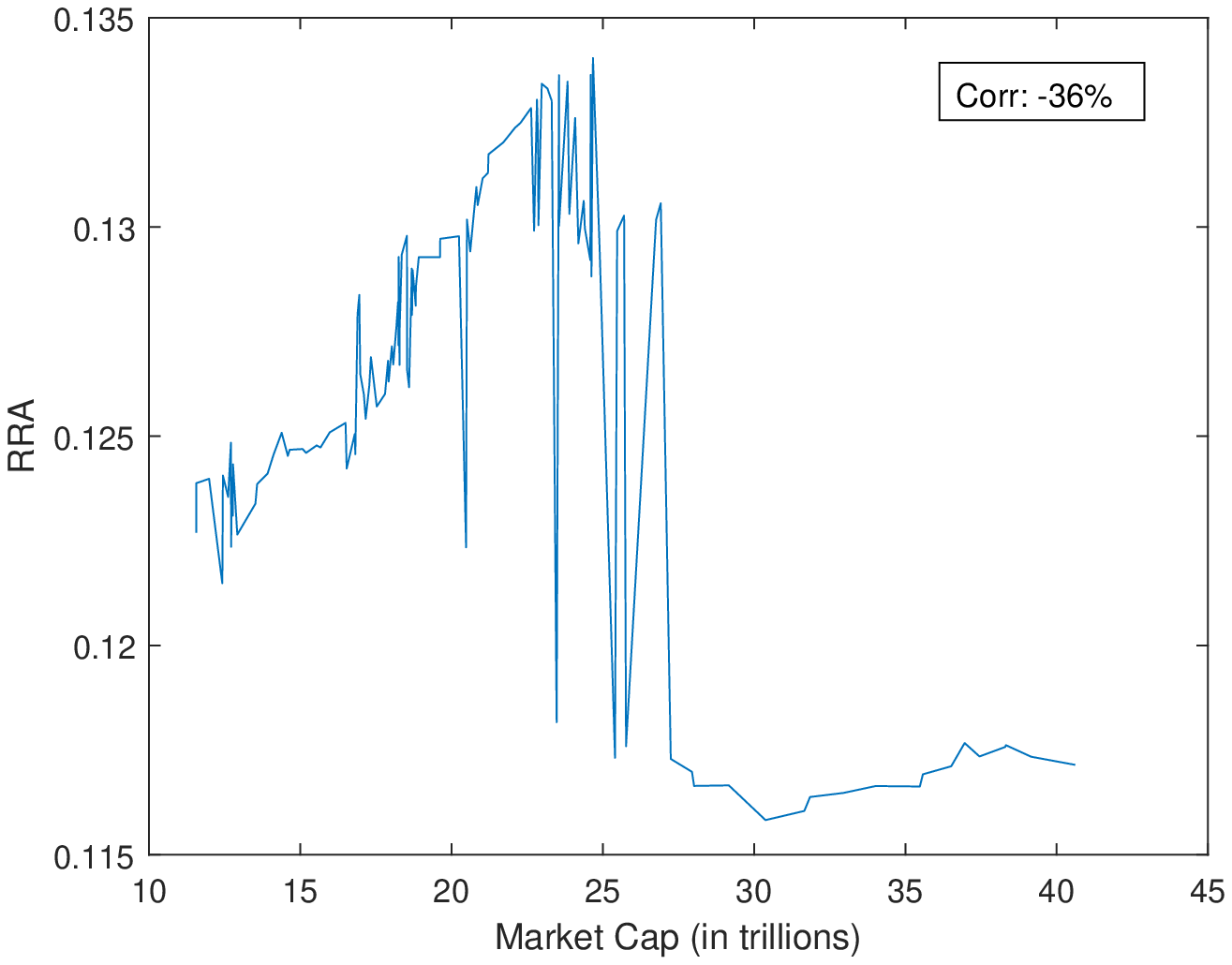}
			\caption{S\&P 500}
		\end{subfigure}
		\caption{RRA with sorted wealth}
	\end{figure}
	
	\newpage
	\begin{figure}[H]
		\begin{subfigure}[b]{0.45\textwidth}	
			\includegraphics[width=\linewidth]{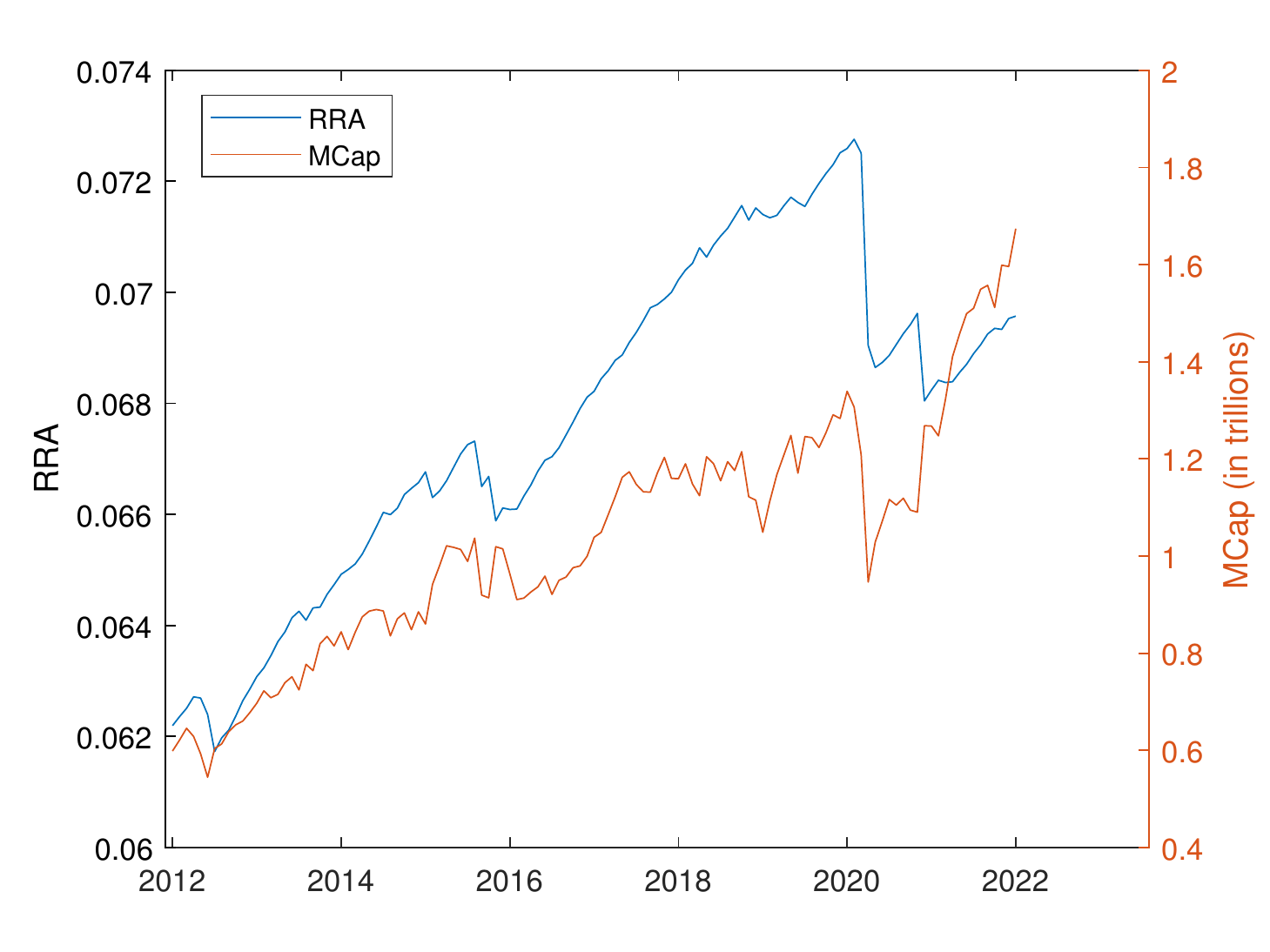}
			\caption{CAC 40}
		\end{subfigure}
		\hspace*{1in}
		\begin{subfigure}[b]{0.45\textwidth}
			\includegraphics[width=\linewidth]{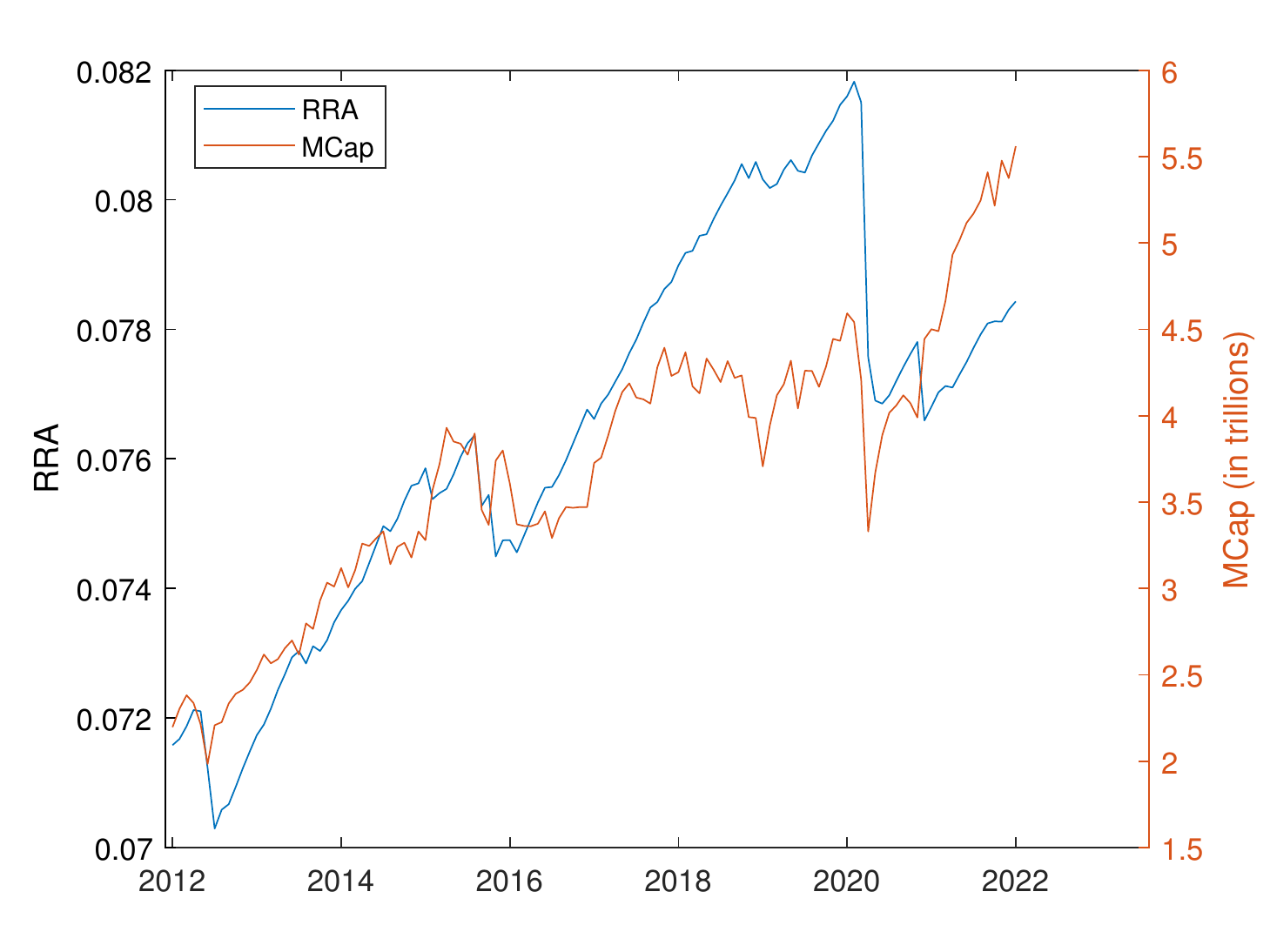}
			\caption{EURO}
		\end{subfigure}
		\hspace*{2in}
		\begin{subfigure}[b]{0.45\textwidth}
			\includegraphics[width=\linewidth]{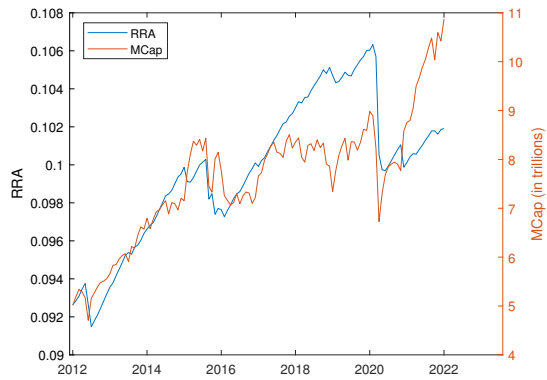}
			\caption{STOXX 600}
		\end{subfigure}
		\caption{RRA in time}
	\end{figure}
	
	\newpage
	\begin{figure}[htp]
		\begin{subfigure}{0.45\textwidth}	
			\includegraphics[width=\linewidth]{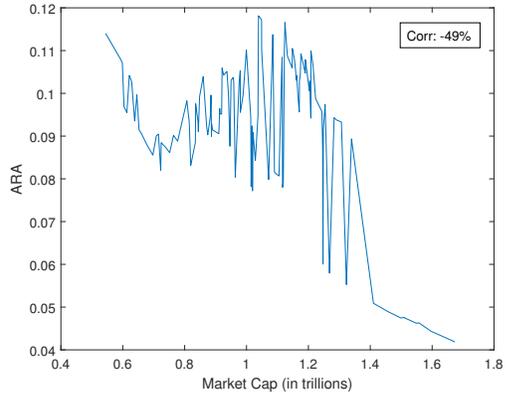}
			\caption{CAC 40}
		\end{subfigure}
		\hspace*{1in}
		\begin{subfigure}{0.45\textwidth}
			\includegraphics[width=\linewidth]{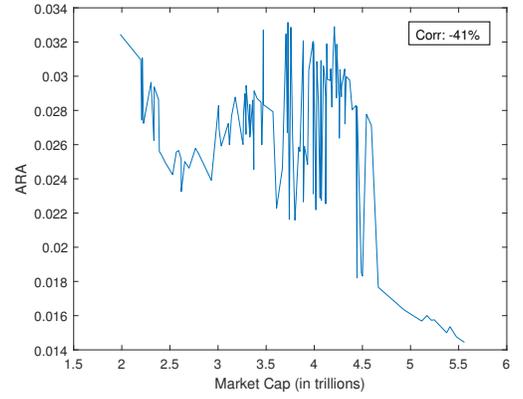}
			\caption{EURO}
		\end{subfigure}
		\begin{subfigure}{0.45\textwidth}
			\includegraphics[width=\linewidth]{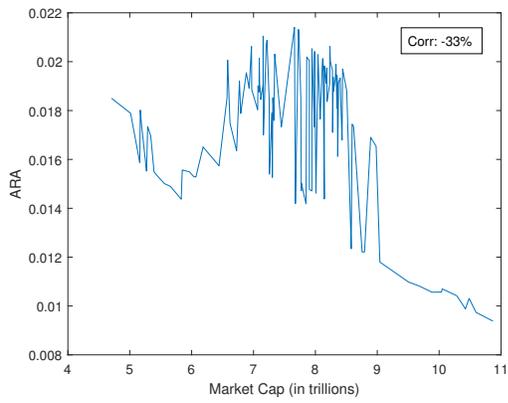}
			\caption{STOXX 600}
		\end{subfigure}
		\hspace*{1in}
		\begin{subfigure}{0.45\textwidth}
			\includegraphics[width=\linewidth]{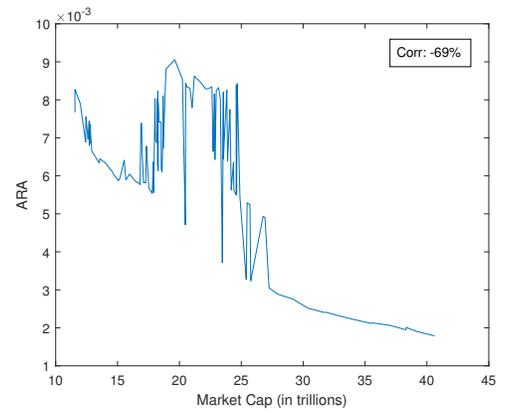}
			\caption{S\&P 500}
		\end{subfigure}
		\caption{ARA with $M=60$}
	\end{figure}
	
	\newpage
	\begin{figure}[htp]
		\begin{subfigure}{0.45\textwidth}	
			\includegraphics[width=\linewidth]{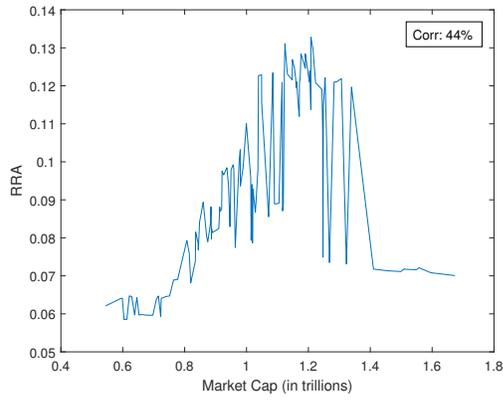}
			\caption{CAC 40}
		\end{subfigure}
		\hspace*{1in}
		\begin{subfigure}{0.45\textwidth}
			\includegraphics[width=\linewidth]{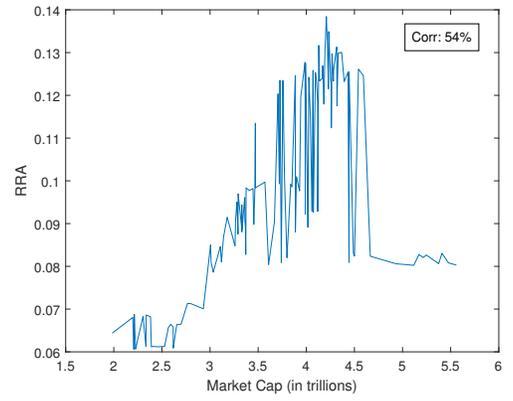}
			\caption{EURO}
		\end{subfigure}
		\begin{subfigure}{0.45\textwidth}
			\includegraphics[width=\linewidth]{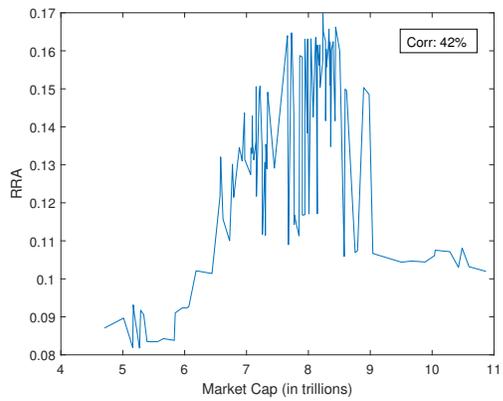}
			\caption{STOXX 600}
		\end{subfigure}
		\hspace*{1in}
		\begin{subfigure}{0.45\textwidth}
			\includegraphics[width=\linewidth]{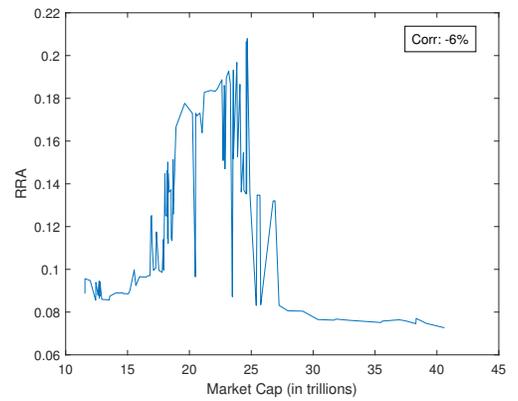}
			\caption{S\&P 500}
		\end{subfigure}
		\caption{RRA with $M=60$}
	\end{figure}
	
	\newpage
	\begin{figure}[htp]
		\begin{subfigure}{0.45\textwidth}	
			\includegraphics[width=\linewidth]{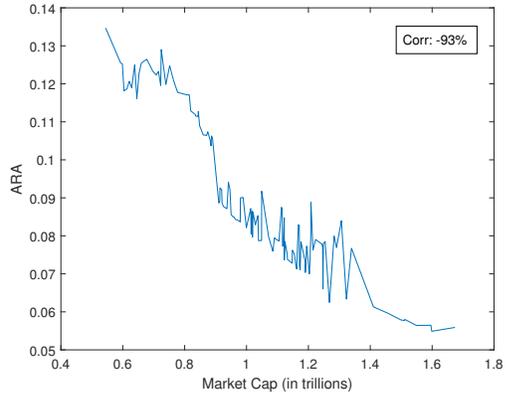}
			\caption{CAC 40}
		\end{subfigure}
		\hspace*{1in}
		\begin{subfigure}{0.45\textwidth}
			\includegraphics[width=\linewidth]{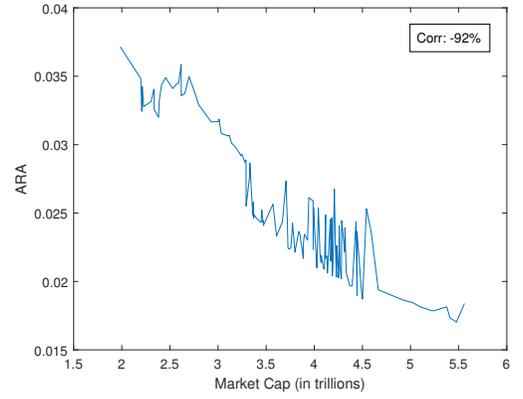}
			\caption{EURO}
		\end{subfigure}
		\begin{subfigure}{0.45\textwidth}
			\includegraphics[width=\linewidth]{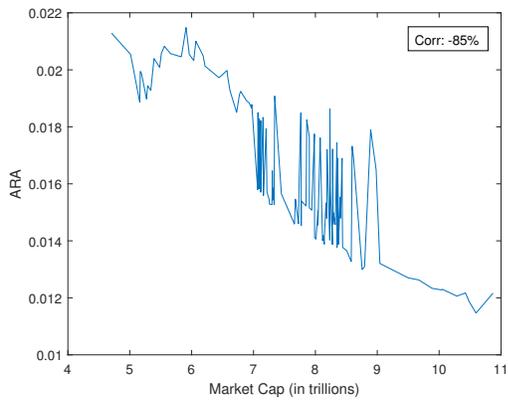}
			\caption{STOXX 600}
		\end{subfigure}
		\hspace*{1in}
		\begin{subfigure}{0.45\textwidth}
			\includegraphics[width=\linewidth]{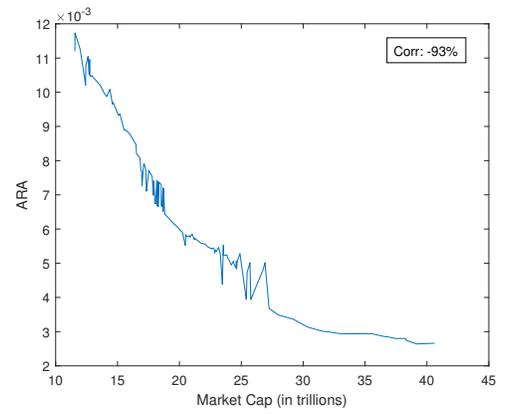}
			\caption{S\&P 500}
		\end{subfigure}
		\caption{ARA with $M=120$}
	\end{figure}
	
	\newpage
	\begin{figure}[H]
		\begin{subfigure}{0.45\textwidth}	
			\includegraphics[width=\linewidth]{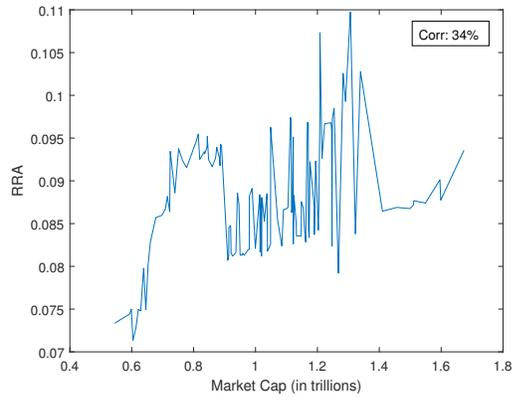}
			\caption{CAC 40}
		\end{subfigure}
		\hspace*{1in}
		\begin{subfigure}{0.45\textwidth}
			\includegraphics[width=\linewidth]{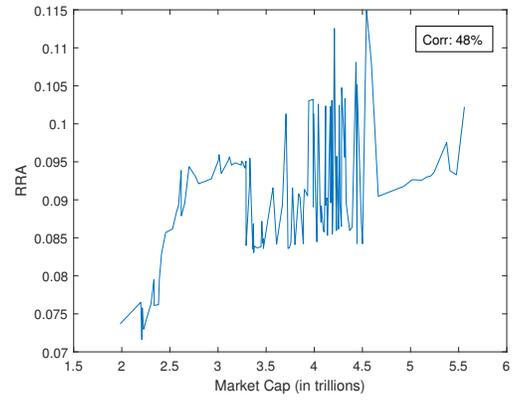}
			\caption{EURO}
		\end{subfigure}
		\begin{subfigure}{0.45\textwidth}
			\includegraphics[width=\linewidth]{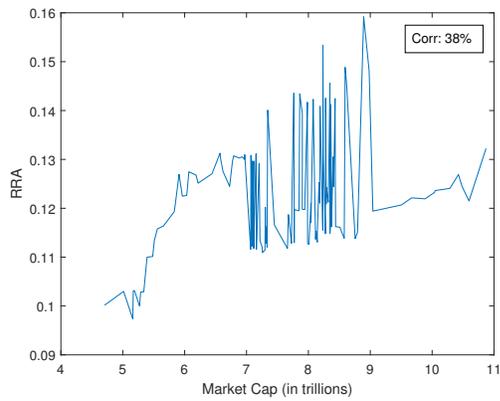}
			\caption{STOXX 600}
		\end{subfigure}
		\hspace*{1in}
		\begin{subfigure}{0.45\textwidth}
			\includegraphics[width=\linewidth]{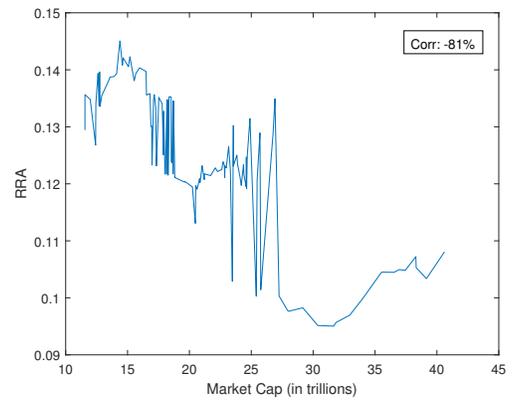}
			\caption{S\&P 500}
		\end{subfigure}
		\caption{RRA with $M=120$}
	\end{figure}
	
	\newpage
	\begin{figure}[htp]
		\begin{subfigure}{0.45\textwidth}	
			\includegraphics[width=\linewidth]{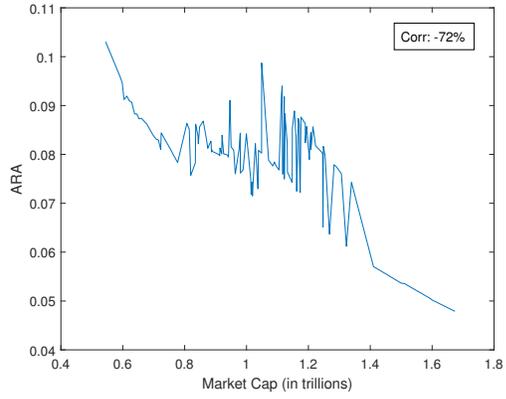}
			\caption{CAC 40}
		\end{subfigure}
		\hspace*{1in}
		\begin{subfigure}{0.45\textwidth}
			\includegraphics[width=\linewidth]{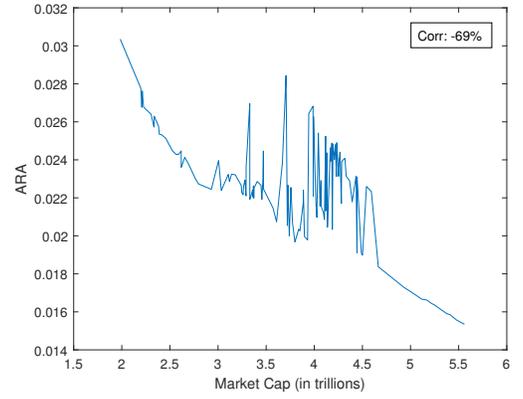}
			\caption{EURO}
		\end{subfigure}
		\begin{subfigure}{0.45\textwidth}
			\includegraphics[width=\linewidth]{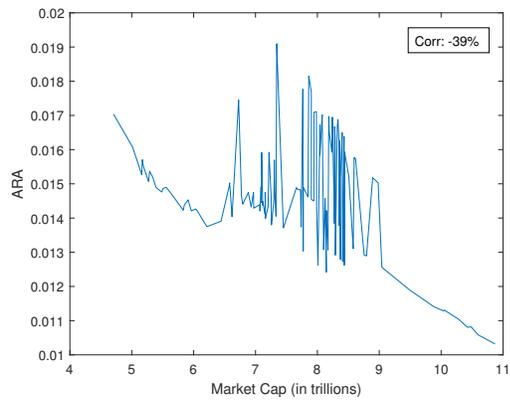}
			\caption{STOXX 600}
		\end{subfigure}
		\hspace*{1in}
		\begin{subfigure}{0.45\textwidth}
			\includegraphics[width=\linewidth]{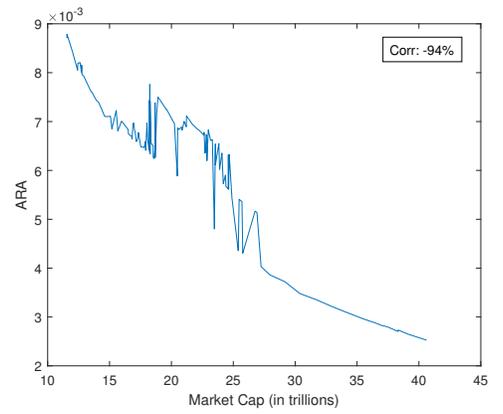}
			\caption{S\&P 500}
		\end{subfigure}
		\caption{ARA with $M=180$}
	\end{figure}
	
	\newpage
	\begin{figure}[H]
		\begin{subfigure}{0.45\textwidth}	
			\includegraphics[width=\linewidth]{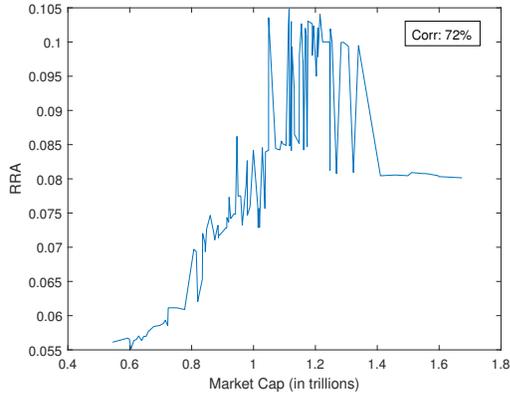}
			\caption{CAC 40}
		\end{subfigure}
		\hspace*{1in}
		\begin{subfigure}{0.45\textwidth}
			\includegraphics[width=\linewidth]{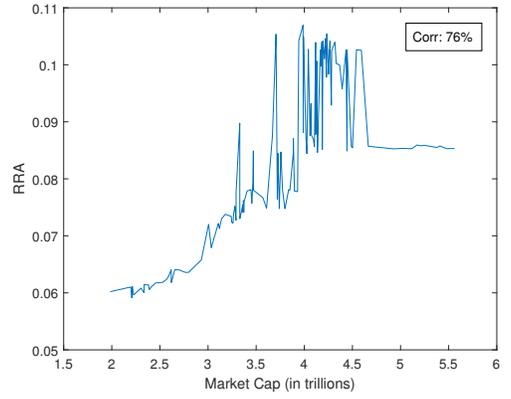}
			\caption{EURO}
		\end{subfigure}
		\begin{subfigure}{0.45\textwidth}
			\includegraphics[width=\linewidth]{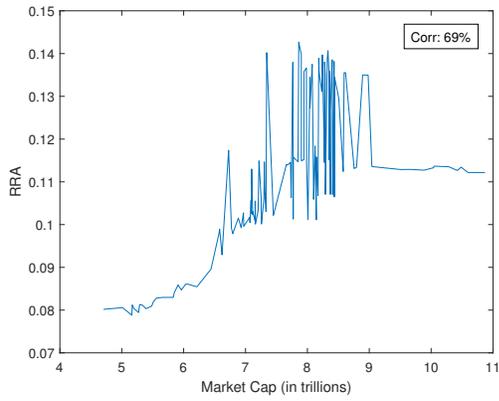}
			\caption{STOXX 600}
		\end{subfigure}
		\hspace*{1in}
		\begin{subfigure}{0.45\textwidth}
			\includegraphics[width=\linewidth]{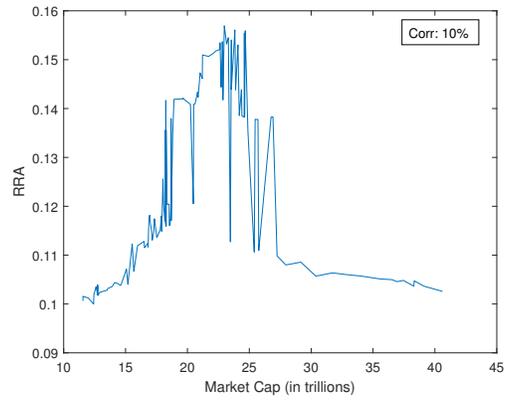}
			\caption{S\&P 500}
		\end{subfigure}
		\caption{RRA with $M=180$}
	\end{figure}

	\newpage
	%%%%%%%%%%%%%%%%%%%%%
	%%%%%%%%%%%%%%%%%%%%%
	%%%Extra bibliography
	%%%%%%%%%%%%%%%%%%%%%
	%%%%%%%%%%%%%%%%%%%%%
	\printbibliography

\end{document}